\documentclass[twocolumn,aps,prd,amssymb,eqsecnum,nofootinbib,superscriptaddress,showpacs]{revtex4}
\usepackage{ifpdf}
\ifpdf
\usepackage{hyperref}
\else
\usepackage[dvipdfm]{hyperref}
\fi

\usepackage{amsmath}
\usepackage{amssymb}
\usepackage{epsfig}
\usepackage{bm}
\usepackage{slashbox}
\usepackage{graphicx}
\usepackage{psfrag}
\usepackage{rotating}
\usepackage{multirow}
\usepackage[percent]{overpic}

\newcommand{\be}{\begin{equation}}
\newcommand{\ee}{\end{equation}}
\newcommand{\ba}{\begin{eqnarray}}
\newcommand{\ea}{\end{eqnarray}}
\newcommand{\bma}{\begin{pmatrix}}
\newcommand{\ema}{\end{pmatrix}}

\newcommand{\Caltech}{\affiliation{Theoretical Astrophysics 350-17,
    California Institute of Technology, Pasadena, California 91125, USA}}
\newcommand{\Cornell}{\affiliation{Center for Radiophysics and Space
    Research, Cornell University, Ithaca, New York 14853, USA}}
\newcommand{\Oberlin}{\affiliation{Department of Physics and Astronomy, Oberlin College, Oberlin, Ohio 44074, USA}}
\newcommand{\GWPAC}{\affiliation{Gravitational Wave Physics and Astronomy Center, California State University Fullerton, Fullerton, California 92831, USA}}

\begin{document}

\title{Visualizing Spacetime Curvature via Frame-Drag Vortexes and Tidal Tendexes \\ II. Stationary Black Holes}

\author{Fan Zhang} \Caltech
\author{Aaron Zimmerman} \Caltech
\author{David A.\ Nichols} \Caltech
\author{Yanbei Chen} \Caltech
\author{Geoffrey Lovelace} \Cornell \GWPAC
\author{Keith D.\ Matthews} \Caltech
\author{Robert Owen} \Cornell \Oberlin
\author{Kip S.\ Thorne} \Caltech 
\date{printed \today}

\date{printed \today}

\begin{abstract}
When one splits spacetime into space plus time, the Weyl curvature tensor
(which equals the Riemann tensor in vacuum) splits 
into two spatial, symmetric, traceless
tensors: the \emph{tidal field} $\boldsymbol{\mathcal E}$, which produces tidal forces,
and the \emph{frame-drag field} $\boldsymbol{\mathcal B}$, which produces differential
frame dragging.
In recent papers, we and colleagues have introduced ways to
visualize these two fields: 
\emph{tidal tendex lines} (integral curves of the three 
eigenvector fields
of $\boldsymbol{\mathcal E}$) and their \emph{tendicities} 
(eigenvalues of these eigenvector fields); and the corresponding entities
for the frame-drag field:
\emph{frame-drag vortex lines} and their \emph{vorticities}.  
These entities fully characterize the vacuum Riemann tensor.
In this paper, we compute and depict the tendex and vortex lines, and their
tendicities and vorticities, outside
the horizons of stationary (Schwarzschild and Kerr) black holes; and we 
introduce and depict the black holes' \emph{horizon tendicity and vorticity} 
(the normal-normal components of $\boldsymbol{\mathcal E}$ and 
$\boldsymbol{\mathcal B}$ on the horizon).  For
Schwarzschild and Kerr black holes, the horizon tendicity is proportional
to the horizon's intrinsic scalar curvature, and the horizon vorticity 
is proportional to an extrinsic scalar curvature.

We show that, for horizon-penetrating time slices,
 all these entities ($\boldsymbol{\mathcal E}$, $\boldsymbol{\mathcal B}$, 
the tendex lines and vortex lines, the lines' tendicities and vorticities, 
and the
horizon tendicities and vorticities) are affected only weakly by changes of
slicing and changes of spatial coordinates, within those slicing and coordinate choices
that are commonly used for black holes. We also explore how the tendex and vortex lines change as the spin of a black hole is increased, and we find, for example, that as a black hole is spun up through a dimensionless spin $a/M = \sqrt3\,/2$, the horizon tendicity at its poles changes sign, 
and an observer hovering or falling inward there switches from being
stretched radially to being squeezed. 
At this spin, the tendex lines that stick out from the horizon's poles switch 
from reaching radially outward toward infinity, to emerging from one pole, 
swinging poloidally around the hole and descending into the other pole.
\end{abstract}


\maketitle

\section{Motivation and Overview}
\label{sec:intro}

It has long been known that, when one performs a 3+1 split of spacetime into
space plus time, the Weyl curvature tensor $C_{\alpha\beta\gamma\delta}$ gets
split into two spatial, symmetric, traceless tensors: the so-called
``electric'' part, $\boldsymbol{\mathcal E}$, 
which we call the \emph{tidal field}
(because it is responsible for the gravitational
stretching and squeezing that generates
tides), and the so-called ``magnetic'' part $\boldsymbol{\mathcal B}$, 
which we call the \emph{frame-drag
field} (because it generates differential frame dragging, i.e., 
differential precession of gyroscopes).  

Recently~\cite{OwenEtAl:2011,Nichols:2011pu}, we and
colleagues have proposed visualizing the tidal field by means of the integral
curves of its three eigenvector fields, which we call \emph{tendex lines},
and each line's eigenvalue, which we call its \emph{tendicity}.  These 
are very much like electric field lines and the magnitude of the electric
field.  Similarly, we have proposed visualizing the frame-drag field by
integral curves of its three eigenvector fields, which we call 
\emph{vortex lines},
and each curve's eigenvalue, which we call its \emph{vorticity}.  
These are analogous to magnetic field lines and the magnitude of the magnetic
field.

In our initial presentation \cite{OwenEtAl:2011}
of these new concepts and their applications,
we demonstrated that they can be powerful tools for
visualizing the nonlinear dynamics of curved spacetime that is triggered
by the inspiral, collision, and merger of binary black holes.  We expect
them also to be powerful visualization tools in other venues of
nonlinear spacetime dynamics (geometrodynamics).

After our initial presentation \cite{OwenEtAl:2011}, 
we have turned to a methodical 
exploration of these tools, in a series of papers in this journal.
We are beginning in Papers I--III by applying these tools to 
``analytically understood''  spacetimes, in order to gain intuition into the 
relation between their visual pictures and the analytics. Then 
in Paper IV and thereafter, we shall apply them to numerical spacetimes, 
looking for types of features we have already found, and retrieving their analytical origin.   

In \cite{Nichols:2011pu} (henceforth Paper I), using examples of nearly flat 
(linearized) spacetimes, 
we have shown that tendex lines and vortex lines can illustrate very well the spacetime dynamics around oscillating multipole sources, and we 
have connected various features of the field lines to physical 
understanding, and to the analytics.  
We found that, in the near zone of an oscillating multipole, the 
field lines are attached to the source; in the transition zone, retardation effects cause the field lines to change 
character in understandable ways; and in the wave zone, the field lines 
approach those of freely propagating plane waves.  In a supplementary 
study~\cite{Zimmerman2011}, some of us have classified the 
tendex and vortex lines of asymptotically flat space times at future null 
infinity according to the lines' topological features. 

Recently, Dennison and Baumgarte \cite{Dennison2012} computed the tendex and 
vortex fields of approximate analytical solutions of boosted, non-spinning 
black holes (both isolated holes and those in binaries).
Specifically, they computed an analytical initial-data solution of the Einstein
constraint equations (in the form of that of Bowen and York 
\cite{Bowen-York:1980}) that is accurate through leading order in a boost-like 
parameter of the black holes.
Their results are an important analytical approximation to the vortex and
tendex fields of a strong-field binary, and will likely be useful for 
understanding aspects of numerical-relativity simulations of binary black 
holes.

In Paper III, we shall explore the tendex and vortex lines, and their tendicities
and vorticities, for quasinormal-mode oscillations of black holes---and shall
see very similar behaviors to those we found, in the linearized approximation,
in Paper I \cite{Nichols:2011pu}.  In preparation for this, we must explore
in depth the application of our new tools to stationary (Schwarzschild and
Kerr) black holes.  That is the purpose of this Paper II.  

In Paper IV we shall apply our tools to numerical simulations of binary-black-hole inspiral,
collision, and merger, and shall use our linearized visualizations (Paper I),
our stationary-black-hole visualizations (Paper II), our 
quasinormal-mode visualizations (Paper III), and Dennison and Baumgarte's 
visualizations \cite{Dennison2012} to gain insight into the
fully nonlinear spacetime dynamics that the binary black holes trigger.

This paper is organized as follows: 
In Sec.~\ref{sec:Review}, we briefly review the underlying theory of the $3+1$ 
split of spacetime and our definitions of the tidal field $\boldsymbol{\mathcal E}$ 
and frame-drag field $\boldsymbol{\mathcal B}$ in~\cite{OwenEtAl:2011,Nichols:2011pu}.
In Sec.~\ref{sec:Horizons}, we introduce the concepts of \emph{horizon
tendicity} (the normal-normal component of $\boldsymbol{\mathcal E}$ on a black-hole
horizon) and \emph{horizon vorticity} (the normal-normal component
of $\boldsymbol{\mathcal B}$), which, for stationary black holes, can be related to the real and imaginary parts of the Newman-Penrose Weyl 
scalar $\Psi_2$ and are the horizon's scalar intrinsic curvature and scalar
extrinsic curvature (aside from simple multiplicative factors).

In Sec.~\ref{sec:StaticSchw}, we give formulae for the 
eigenvector and eigenvalue 
fields for the tidal field around a static (Schwarzschild) black hole, 
we draw pictures
of the black hole's corresponding tendex lines, and we discuss the connection
to the tidal stretching and squeezing felt by observers near a Schwarzschild
hole. (The frame-drag field vanishes for a Schwarzschild hole.)

In Sec.~\ref{sec:SlowKerr}, we turn on a slow rotation of the hole, we 
compute the frame-drag field $\boldsymbol{\mathcal B}$
generated by that rotation, we visualize $\boldsymbol{\mathcal B}$ via 
color-coded pictures of the horizon vorticity and the vortex lines,
and we discover a spiraling of azimuthal tendex lines that is created
by the hole's rotation.  In this section, we restrict ourselves to 
time slices (and the fields on those time slices) that have constant
ingoing Eddington-Finklestein time, and that therefore penetrate the
horizon smoothly.  (For the Schwarzschild black hole of 
Sec.~\ref{sec:StaticSchw}, the tendex lines are the same in Schwarzschild
slicing as in Eddington-Finklestein slicing; the hole's rotation destroys this.)

In Sec.~\ref{sec:StaticKerr}, we turn to rapidly rotating (Kerr) black holes,
and explore how the vortex and tendex lines and the horizon vorticities
and tendicities change when a hole is spun up to near maximal angular
velocity.
In these explorations, we restrict ourselves to horizon-penetrating slices, 
specifically: slices of constant Kerr-Schild time $\tilde t$, and the 
significantly different slices of constant Cook-Scheel, harmonic time
$\bar t$.
By using the same spatial coordinates in the two cases, we
explore how the time slicing affects the tendex and vortex lines and
the horizon tendicities and vorticities.  There is surprisingly little
difference, for the two slicings; the field lines and horizon properties
change by only modest amounts when one switches from one slicing to the
other (top row of Fig.~\ref{fig:CompareSlicing} 
compared with
bottom row).  By contrast, when we use non-horizon-penetrating Boyer-Lindquist slices (Appendix \ref{app:BLKerr}), the field lines are
noticeably changed. In Sec.~\ref{sec:StaticKerr} we also explore how
the vortex and tendex lines (plotted on a flat computer screen or
flat sheet of paper) change, when we change the spatial coordinates
with fixed slicing (Fig.~\ref{fig:CompareCoords}).  We find only modest changes, and they
are easily understood and quite obvious once one understands the relationship
between the spatial coordinate systems.
 
In Sec.~\ref{Conclusion}, we briefly summarize our results. 
In three Appendices, we present mathematical details that underlie some of
the results in the body of the paper.

Throughout this paper we use geometrized units, with $G = c =1$. 
Greek indices are used for 4D spacetime quantities, and run from 0 to 3. 
Latin indices are used for spatial quantities, and run from 1 to 3. 
Hatted indices indicate components on an orthonormal tetrad. 
Capital Latin indices from the start of the alphabet are used 
for angular quantities defined on spheres of some constant radius, 
and they generally run over $\theta, \phi$. 
We use signature $(-+++)$ for the spacetime metric, and our Newman-Penrose quantities are defined appropriately for this signature, as in~\cite{Stephani2003}.

\section{Tendex and Vortex Lines}
\label{sec:Review}

In this section we will briefly review the $3+1$ split and the definition of 
our spatial curvature quantities. 
A more detailed account is given in Paper I of this 
series~\cite{Nichols:2011pu}. 
To begin with, we split the spacetime using a unit timelike vector $\vec u$, 
which is everywhere normal to the slice of constant time. 
This vector can be associated with a family of observers who travel with 
four-velocity $\vec u$, and will observe the corresponding time slices as 
moments of simultaneity. 
We consider only vacuum spacetimes, where the Riemann tensor 
$R_{\mu \nu \rho \sigma}$ is the same as the Weyl tensor 
$C_{\mu \nu \rho \sigma}$. 
The Weyl tensor has ten independent degrees of freedom, and these are encoded 
in two symmetric, traceless spatial tensors $\boldsymbol{\mathcal E}$ and 
$\boldsymbol{\mathcal B}$. 
These spatial tensors are formed by projection of the Weyl tensor 
$C_{\mu \nu \rho \sigma}$ and (minus) its Hodge dual 
$^*C_{\mu \nu \rho \sigma}$ onto the spatial slices using $u^\mu$, and the 
spatial projection operator $\gamma_\alpha{}^\mu$. 
The projection operator is given by raising one index on the spatial metric of 
the slice, $\gamma_{\mu \nu} = g_{\mu \nu} + u_\mu u_\nu$. 
The resulting spatial projection of the Weyl tensor is given by an even-parity 
field called the ``electric'' part of $C_{\mu \nu \rho\sigma}$ and also called 
the ``tidal field'',
\begin{subequations}
\label{eq:DefsEijBij}
\begin{equation}  \label{eq:DefEij}
\mathcal E_{\alpha\beta} = {\gamma_\alpha}^\rho {\gamma_\beta}^\sigma 
C_{\rho\mu\sigma\nu}u^\mu u^\nu\;, \quad \hbox{i.e. }
\mathcal E_{ij} = C_{i \hat 0 j \hat 0}\;, 
\end{equation}
and an odd-parity field called the ``magnetic'' part of 
$C_{\mu \nu \rho\sigma}$ and also called the ``frame-drag field'',
\begin{equation}  \label{eq:DefBij}
\mathcal B_{\alpha\beta} = - {\gamma_\alpha}^\rho {\gamma_\beta}^\sigma
\,^*C_{\rho\mu\sigma\nu}u^\mu u^\nu\;, \quad \hbox{i.e. }
 \mathcal B_{ij} = \frac{1}{2} \epsilon_{ipq}C^{pq}_{\phantom{ab}j\hat 0}.
\end{equation}
\end{subequations}
Here, as usual, we give spatial (Latin) indices to quantities after projection 
onto the spatial slices using $\gamma_\alpha{}^\mu$. 
We note that our conventions on the antisymmetric tensors are, when expressed 
in an orthonormal basis, $\epsilon_{\hat 0 \hat 1 \hat 2 \hat 3} = +1$ and 
$\epsilon_{\hat 1 \hat 2 \hat 3} = +1$, with 
$\epsilon_{ijk} = \epsilon_{\hat 0 ijk}$.

The real, symmetric matrices, $\mathcal E_{ij}$ and $\mathcal B_{ij}$ are 
completely characterized by their orthogonal eigenvectors and corresponding 
eigenvalues. 
Note that, since each tensor is traceless, the sum of its three eigenvalues 
must vanish. 
Our program for generating field lines to visualize the spacetime curvature is 
to find these eigenvector fields by solving the eigenvalue problem,
\be
\label{eq:GeneralEigen}
\mathcal E^i{}_j v^j = \lambda v^i \,.
\ee
This results in three eigenvector fields for each of the two tensors 
$\boldsymbol{\mathcal E}$ and $\boldsymbol{\mathcal B}$. 
These fields are vector fields on the spatial slice, and behave as usual under 
transformations of the spatial coordinates (but not changes of the slicing 
vector $u^\mu$). 
By integrating the streamlines of these eigenvector fields, we arrive at a set 
of three {\it tendex} lines and three {\it vortex} lines. 
These lines are associated with the corresponding eigenvalues, the 
{\it tendicity} of each tendex line and {\it vorticity} of each vortex line.  
In visualizations, we color code each tendex or vortex line by its tendicity 
or vorticity. 

This method of visualization represents physical information about the 
spacetime in a very natural way. 
It was shown in Paper I that the tidal field $\boldsymbol{\mathcal E}$ 
describes the local tidal forces between nearby points in the spacetime, and 
the less-familiar frame-drag field $\boldsymbol{\mathcal B}$ describes the 
relative precession of nearby gyroscopes. 
In the local Lorentz frame of two freely falling observers, separated by a 
spatial vector $\xi^j$, the differential acceleration experienced by the 
observers is 
\begin{subequations}
\be
\label{eq:Diffacceleration}
\Delta a^i = - \mathcal E^i{}_j \xi^j \,.
\ee
If these same observers carry inertial guidance gyroscopes, each will measure 
the gyroscope of the other to precess (relative to her own) with a vectorial 
angular velocity dictated by $\boldsymbol{\mathcal B}$,
\be
\label{eq:Diffprecession}
\Delta \Omega^i = \mathcal B^i{}_j \xi^j \,.
\ee
\end{subequations}
In particular, note that if one observer measures a clockwise precession of the
other observer's gyroscope, the second observer will also measure the 
precession of the first to be clockwise.

The physical meaning of the tendex and vortex lines is then clear: if two 
observers have a small separation along a tendex line, they experience an 
acceleration along that line with a magnitude (and sign, in the sense of being 
pushed together or pulled apart) given by the value of the tendicity of that 
line, as governed by Eqs.~\eqref{eq:Diffacceleration}
and~\eqref{eq:GeneralEigen}. 
In the same way, two observers separated along a vortex line experience 
differential frame dragging as dictated by Eqs.~\eqref{eq:Diffprecession} 
and~\eqref{eq:GeneralEigen} 
(with $\mathcal E_{ij} \rightarrow \mathcal B_{ij}$).

\section{Black-Hole Horizons; The Horizon Tendicity $\mathcal E_{NN}$ and Vorticity $\mathcal B_{NN}$}
\label{sec:Horizons}

In many problems of physical interest, such as black-hole perturbations and 
numerical-relativity simulations using excision (as in the {\tt SpEC} 
code~\cite{SpECwebsite}), black-hole interiors are not included in the 
solution domain. 
However, we are interested in structures defined on spacelike surfaces that 
penetrate the horizon, and, in order to retain the information describing the 
dynamics of spacetime in and near the black-hole region, we must define 
quasilocal quantities representing the tendicity and vorticity of the excised 
black-hole region.

We define the horizon tendicity and vorticity as follows:
For a hypersurface-normal observer with 4-velocity $\vec u$, passing through a 
worldtube such as an event horizon or a dynamical horizon, the worldtube has 
an inward pointing normal $\vec N$ orthogonal to $\vec u$, and two orthonormal
vectors tangent to its surface, ${\vec e_2}$ and ${\vec e_3}$ (together these 
four vectors form an orthonormal tetrad).
The horizon tendicity is defined as $\mathcal E_{NN} = \mathcal E_{ij} N^i N^j$
and the horizon vorticity is $\mathcal B_{NN} = \mathcal B_{ij} N^i N^j$.
Physically, they represent the differential acceleration and differential
precession of gyroscopes, respectively, as measured by the observer, for two 
points separated in the direction of ${\vec N}$, and projected along that 
direction.

The horizon tendicity and vorticity have several interesting connections 
with other geometric quantities of 2-surfaces.
In particular, they fit nicely into the Newman-Penrose (NP) formalism
\cite{Newman1962}.
Rather than describe spacetime in terms of the tetrad $\vec u$, ${\vec N}$, 
${\vec e_2}$ and ${\vec e_3}$, the NP approach describes spacetime in terms of
a null tetrad, with two null vectors ${\vec l}$, and ${\vec n}$, together with 
a complex spatial vector ${\vec m}$ and its complex conjugate ${\vec m}^*$.
It is convenient to adapt this tetrad to the 2-surface so that it is given by
\begin{align}
\label{eq:NulltoOrtho}
{\vec l} &= \frac{1}{\sqrt{2}}({\vec u} - {\vec N})\,, &
{\vec n} &=\frac{1}{\sqrt{2}} ({\vec u} + {\vec N}) \,, \notag \\
{\vec  m} &= \frac{1}{\sqrt{2}}({\vec e_2} + i {\vec e_3}) \,.
\end{align} 
On an event horizon, ${\vec l}$ is tangent to the generators of the horizon 
and ${\vec n}$ is the ingoing null normal.
It is not difficult to show that in this tetrad the complex Weyl scalar 
$\Psi_2$ is given by
\be
\Psi_2 = C_{lmm^*n} = (\mathcal E_{NN} + i\mathcal B_{NN})/2 \,,
\ee
where $C_{lmm^*n}$ is the Weyl tensor contracted into the four different null
vectors of the tetrad in the order of the indices.

Penrose and Rindler \cite{Penrose1992} relate the NP quantities to curvature 
scalars of a spacelike 2-surface in spacetime.
In turn, we can then connect their results to the horizon tendicity and
vorticity.
More specifically: Penrose and Rindler define a complex curvature of a 
2-surface that equals 
\be
\mathcal K = \frac 14 \left(\mathcal R + i\mathcal X \right) \,.
\ee
Here $\mathcal R$ is the intrinsic Ricci curvature scalar of a the 2D horizon
and $\mathcal X$ is a scalar extrinsic curvature (a curvature of the
bundle of vector spaces normal to the two-surface in spacetime).
This extrinsic curvature $\mathcal X$ is related to the H\'aj\'i\v{c}ek field
\cite{Damour1982} $\Omega_A = n^\mu \nabla_A l_\mu$ (where $\nabla_A$ 
denotes the covariant derivative projected into the 2D horizon) by
$\mathcal X = \epsilon^{AB} \nabla_A \Omega_B$, where 
$\epsilon^{AB}$ is the antisymmetric tensor of the 2D horizon. In the language of differential forms, $\mathcal X$ is the dual of the exterior derivative of the H\'aj\'i\v{c}ek 1-form.  

Penrose and Rindler \cite{Penrose1992} show that for a general, possibly
dynamical black hole,  
\be
\mathcal K = -\Psi_2 +\mu\rho - \lambda\sigma \,,
\ee
where $\rho$, $\sigma$, $\mu$, and $\lambda$ are spin coefficients related to 
the expansion and shear of the null vectors $\vec l$ and $\vec n$, respectively.
This means that the horizon tendicity and vorticity are given by
\begin{subequations}
\ba
\mathcal E_{NN} & =& -\mathcal R/2 + 2\Re[\mu\rho - \lambda\sigma]\,, \\
\mathcal B_{NN} & =& -\mathcal X/2 + 2\Im[\mu\rho - \lambda\sigma]\,.
\ea
\end{subequations}
In the limit of a stationary black hole (this paper),
$\rho$ and $\sigma$ vanish, so
\be
\mathcal E_{NN} = -\mathcal R/2\;, \quad {\rm and} \quad 
\mathcal B_{NN} = -\mathcal X/2\;.
\ee
The 2D horizon of a stationary black hole has spherical topology, and the
Gauss-Bonnet theorem requires that the integral of the scalar curvature 
$\mathcal R$ over a spherical surface is $8\pi$; accounting for factors of two,
the integral of the horizon tendicity $\mathcal E_{NN}$ over the horizon is 
$-4\pi$ (the average value of the horizon tendicity will be negative). 
Stokes's theorem states that the integral of an exact form such as $\mathcal X$ 
vanishes on a surface of spherical topology, and the horizon vorticity will 
also have zero average.  
In formulae:
\begin{equation}
\oint \mathcal E_{NN} dA  = - 4\pi\,, \quad \oint B_{NN} dA = 0
\label{eq:StationaryHorizonIntegrals}
\end{equation}
\emph{for the horizon of a stationary black hole.}

It is worth noting a few other examples in the literature where the complex 
curvature quantities (and as such, horizon tendicity and vorticity) have been 
used.  
The most common use of horizon vorticity (in a disguised form) is for computing 
the spin angular momentum associated with a quasilocal black-hole horizon.  
Following Refs.~\cite{BrownYork1993, Ashtekar2001, Ashtekar2003}, it has become
common to compute black hole spin in numerical-relativity simulations
using the following integral over the horizon:
\begin{equation}
J = -\frac{1}{8 \pi}\oint K_{ij} N^i \varphi^j dA,
\end{equation}
where $K_{ij}$ is the extrinsic curvature of the spatial slice embedded in 
spacetime, $\vec N$ is the inward-pointing unit normal vector to the horizon
in the spatial slice, and $\vec \varphi$ is a rotation-generating vector 
field tangent to the two-dimensional horizon surface.  If $\vec \varphi$ is 
a Killing vector, then one can show that $J$ is conserved.  In 
Ref.~\cite{Dreyer2003}, this was applied to binary-black-hole simulations 
with $\vec \varphi$ given as a certain kind of approximate Killing vector 
that can be computed even on a deformed two-surface.  In 
Ref.~\cite{Cook2007}, and independently in Refs.~\cite{OwenThesis, 
Lovelace2008}, this idea was refined.  The quantity $J$ can be shown to 
be {\em boost invariant} (independent of boosts of the spatial slice in the 
direction of $\vec N$) if $\vec \varphi$ is divergence-free.  Hence, in 
Refs.~\cite{Cook2007, OwenThesis, Lovelace2008}, $\vec \varphi$ 
is restricted to have the form $\varphi^A = \epsilon^{AB} \nabla_B \zeta$, 
where $\zeta$ is some scalar quantity on the two-surface (eventually fixed 
by a minimization problem for other components of the Killing equation).  
Once this substitution has been made, an integration by parts allows $J$ 
to be written as:
\begin{equation}
J = \frac{1}{8 \pi} \oint {\mathcal X} \zeta dA.
\end{equation}
The quantity $\zeta$ is 
fixed by a certain eigenvalue problem on the horizon 2-surface.  On a round 2-sphere, the operator 
in this eigenproblem reduces to the conventional Laplacian, 
and $\zeta$ can be shown to reduce to an $\ell = 1$ spherical 
harmonic. Thus the quasilocal black-hole spin defined in 
Refs.~\cite{Cook2007, OwenThesis, Lovelace2008} can be thought 
of as the dipole part of the horizon vorticity. 

There are simpler ways that one can distill a measure of black hole spin 
from the concepts of horizon vorticity and tendicity. In 
Ref.~\cite{Lovelace2008}, an alternative measure of spin was made by 
comparing the maximum and minimum values of the horizon scalar curvature 
to formulae for a Kerr black hole. This method has roots in older techniques 
by which spin is inferred from the horizon's intrinsic geometry through 
measurements of geodesic path length (see, for example, 
Refs.~\cite{AnninosEtAl:1994, BrandtSeidel:1995, AlcubierreEtAl:2005}). 
The method of computing spin by comparing 
horizon curvature extrema to Kerr formulae could be extended to use the 
extrinsic scalar curvature, or the horizon vorticity or tendicity (which 
differ from the scalar curvatures in dynamical situations). While such methods 
have the benefit of relative simplicity, their practical value in numerical 
relativity is weakened by an empirical sensitivity of the inferred spin to 
effects such as ``junk'' radiation and black hole 
tides~\cite{Lovelace2008, Chu2009}.

In Ref.~\cite{Ashtekar2004}, it was shown that higher spherical-harmonic components of these horizon quantities provide natural definitions of {\em source multipoles} on 
axisymmetric isolated horizons.  In Refs.~\cite{Schnetter2006} 
and~\cite{Owen2009}, this formalism 
was extended to less symmetric cases for use with numerical-relativity 
simulations, while attempting to introduce as little gauge ambiguity as 
possible in the 
process.  Related applications of this formalism can be found in 
Refs.~\cite{Vasset2009, Jasiulek2009}.

\section{Schwarzschild Black Hole}
\label{sec:StaticSchw}

In this section, we examine vortex and tendex lines for a non-rotating black 
hole with mass $M$.  These lines, of course, depend on our choice of time slicing.
As in the numerical simulations that are the focus of Paper IV, so also
here, we shall use a slicing that penetrates smoothly through the black hole's
horizon.  The slices of constant Schwarzschild time $t$ for the hole's 
Schwarzschild metric
\be
ds^2 = - (1-2M/r) dt^2 + {dr^2 \over 1-2M/r} + r^2 d\theta^2 
+ r^2 \sin^2\theta d\phi^2
\label{eq:SchwarzschildMetric}
\ee 
do not penetrate the horizon smoothly; rather, they become singular 
as they approach the horizon.  
(Dennison and Baumgarte \cite{Dennison2012} compute the tidal and frame-drag 
fields of a Schwarzschild black hole in a slice of constant Schwarzschild time 
and in isotropic coordinates; see their paper for comparison.)

The simplest horizon-penetrating slices are those of constant ingoing 
Eddington-Finkelstein (EF) time 
\be 
\tilde t = t+ 2M \ln|r/2M -1|\;.
\label{eq:tEFdef}
\ee
The Schwarzschild metric (\ref{eq:SchwarzschildMetric}), rewritten using
EF coordinates $\{\tilde t, r, \theta, \phi\}$, takes the form
\ba
ds^2 &=& -\biggl( 1 - \frac{2M}{r}\biggl) d \tilde t^2 + \frac{4M}{r} d\tilde t dr + \biggl( 1 + \frac{2M}{r}\biggl) dr^2 \nonumber\\ 
&&+ r^2 d\theta^2 + r^2\sin^2\theta d\phi^2\;.
\label{eq:EFMetric}
\ea
The observers who measure the tidal and frame-drag fields that lie in 
a slice of constant $\tilde t$ have 4-velocities
$\vec u = - \alpha_{\rm EF} \vec\nabla \tilde t$, where $\alpha_{\rm EF}
= 1/\sqrt{1+2M/r}$ is
the normalizing lapse function. These observers can be regarded as carrying
the following orthonormal tetrad for use in their measurements:
\ba
\vec{u} & =& \frac{1} {\sqrt{1+ 2M/r}}\biggl[ \biggl(1+\frac{2M}{r} \biggr) \partial_{\tilde t} - \frac{2M}{r} \partial_r \biggr]  \,, \notag \\
\vec{e}_{\hat r} & = & \frac{1} {\sqrt{1+ 2M/r}} \partial_r \,, \qquad \qquad
\vec{e}_{\hat \theta} =  \frac{1}{r} \partial_\theta\,, \notag \\
 \vec{e}_{\hat \phi} & =&  \frac{1}{r \sin \theta} \partial_\phi \,.
\label{eq:EFtetrad}
\ea
The nonzero components of the tidal field that they measure using this 
tetrad are
\be
\mathcal E_{\hat r \hat r} = -{2M\over r^3}\;, \quad 
\mathcal E_{\hat\theta \hat\theta} = \mathcal E_{\hat\phi \hat\phi} = 
{M\over r^3}\;,
\label{eq:EFtidal}
\ee
and the frame-drag field $\mathcal B_{\hat a \hat b}$ vanishes. 
(See, e.g., Eq.~(31.4b) of \cite{MTW}). 

Note that the black hole's tidal field (\ref{eq:EFtidal}) has the same form as
the Newtonian tidal tensor outside of a spherical source.  
Since the tidal field is diagonal in this tetrad, its eigenvalues and its 
unit-normed eigenvectors are  
\begin{align}
\vec{V_r} & = \vec{e}_{\hat r} \,, & \lambda_r &= - \frac{2M}{r^3} \,, \notag \\
\vec{V_\theta} &= \vec{e}_{\hat \theta} \,, & \lambda_\theta &= \frac{M}{r^3} \,, \notag \\
\vec{V_\phi} &= \vec{e}_{\hat \phi} \,, & \lambda_\phi &= \frac{M}{r^3} \,.
\label{eq:SchTendicities}
\end{align}
Because the two transverse eigenvalues $\lambda_\theta$ and $\lambda_\phi$ are 
degenerate, any vector in the transverse vector space spanned by 
$\vec e_{\hat\theta}$ and $\vec e_{\hat\phi}$ is a solution to the eigenvalue 
problem, and correspondingly, any curve that lies in a sphere of constant $r$ 
can be regarded as a tendex line. 
However (as we shall see in the next section), when the black hole is given
an arbitrarily small rotation about its polar axis $\theta=0$, the 
degeneracy is broken, the non-degenerate transverse eigenvectors become
$\vec e_{\hat\theta}$ and $\vec e_{\hat\phi}$, and the transverse 
tendex lines become circles of constant latitude and longitude. 

\begin{figure}[t]
\includegraphics[width=0.57\columnwidth]{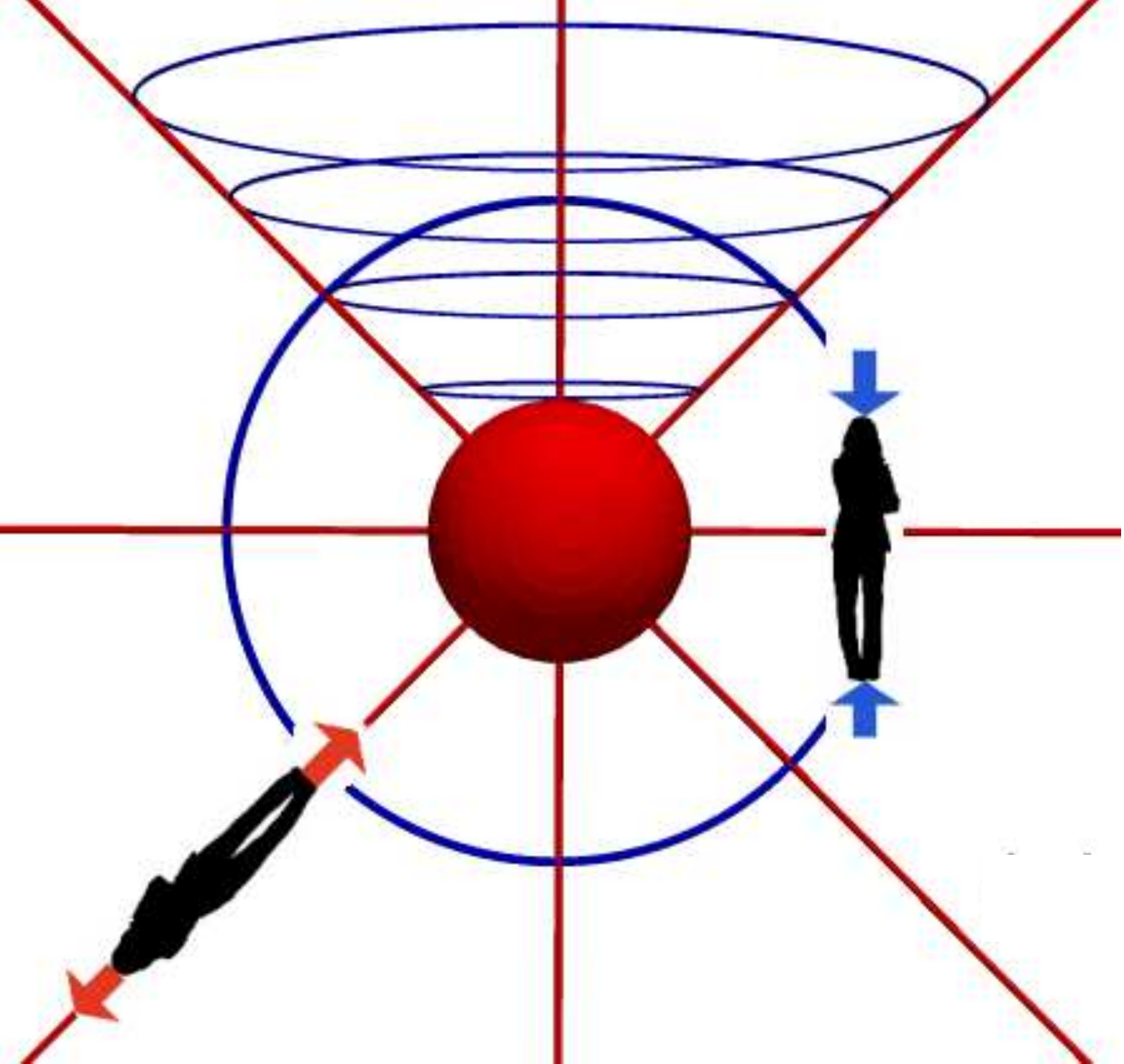}
\caption{
Tendex lines for a non-rotating (Schwarzschild) black hole. These lines 
are identical to those generated by a spherically symmetric mass distribution in the Newtonian limit. Also shown are observers who experience the tidal stretching and compression associated with the tendex lines.
}
\label{fig:SchwTendex}
\end{figure}

In Figure~\ref{fig:SchwTendex}, we plot a few of these transverse tendex
lines (giving them a blue color corresponding to positive tendicity
$\lambda_\theta >0$ and $\lambda_\phi>0$), 
and also a few of the radial tendex lines (colored red for negative
tendicity $\lambda_r < 0$). 
Also shown are two human observers, one oriented along
a blue tendex line (and therefore being squeezed by the tidal field) the
other oriented along a red tendex line (and therefore being stretched). 

\section{Slowly Rotating Black Hole}
\label{sec:SlowKerr}

\subsection{Slicing and coordinates}
\label{subsec:SlowKerrSlicing}

When the black hole is given a slow rotation with angular momentum per unit
mass $a$, its metric (\ref{eq:SchwarzschildMetric}) in Schwarzschild 
coordinates acquires an off-diagonal $g_{t\phi}$ term:
\ba
ds^2 &=& - (1-{2M/ r}) dt^2 + {dr^2 \over 1-2M/r} 
+ r^2 d\theta^2 \nonumber\\
&& + r^2 \sin^2\theta d\phi^2 - {4 a M \over r}\sin^2\theta dt d\phi\;
\label{eq:SlowKerr}
\ea
[the Kerr metric in
Boyer-Lindquist coordinates, Eq.\ (\ref{eq:BLLineElem2}) below, 
linearized in $a$].  The
slices of constant EF time $\tilde t = t + 2M \ln |r/2M -1|$ are still
smoothly horizon penetrating, but the dragging of inertial frames (the
off-diagonal $g_{t\phi}$ term in the metric) causes the Schwarzschild
$\phi$ coordinate to become singular at the horizon.  To fix this, we must
``unwrap'' $\phi$, e.g., by switching to 
\be
\tilde \phi = \phi + (a/ 2M) \ln|1-2M/r|\;,
\label{eq:EFphiSlow}
\ee
thereby bringing the ``slow-Kerr'' metric (\ref{eq:SlowKerr}) into the form
\ba
ds^2 &=& -\biggl( 1 - \frac{2M}{r}\biggl) d \tilde t^2 + \frac{4M}{r} d\tilde t dr + \biggl( 1 + \frac{2M}{r}\biggl) dr^2 \nonumber\\
&&+ r^2 d\theta^2 + r^2\sin^2\theta d\tilde\phi^2 
-{4aM\over r} \sin^2\theta d\tilde td\tilde\phi \nonumber\\
&&- 2a\sqrt{1+2M/r}\sin^2\theta
dr d\tilde\phi \;
\label{eq:SlowKerrMetric}
\ea
[Eq.\ (\ref{KerrIngoingMetric}) below, linearized in $a$], which is well
behaved at and through the horizon.  The observers who move orthogonally
to the slices of constant $\tilde t$ have 4-velocity $\vec u$ and 
orthonormal basis the same as for a non-rotating black hole, Eq.\ 
(\ref{eq:EFtetrad}), except that $\vec e_{\hat r}$ is changed to
\be
\vec e_{\hat r} = {1\over \sqrt{1+2M/r}} \left[ \partial_r 
+ {a\over r^2}(1+2M/r)\partial_{\tilde\phi} \right]
\label{eq:tetradSlowRot}
\ee
[Eq.\ (\ref{eq:KerrIKtetrad}) below, linearized in $a$].

\subsection{Frame-drag field and deformed tendex lines}
\label{subsec:SlowKerrFD}

\begin{figure*}
\begin{overpic}[width=0.9\columnwidth]{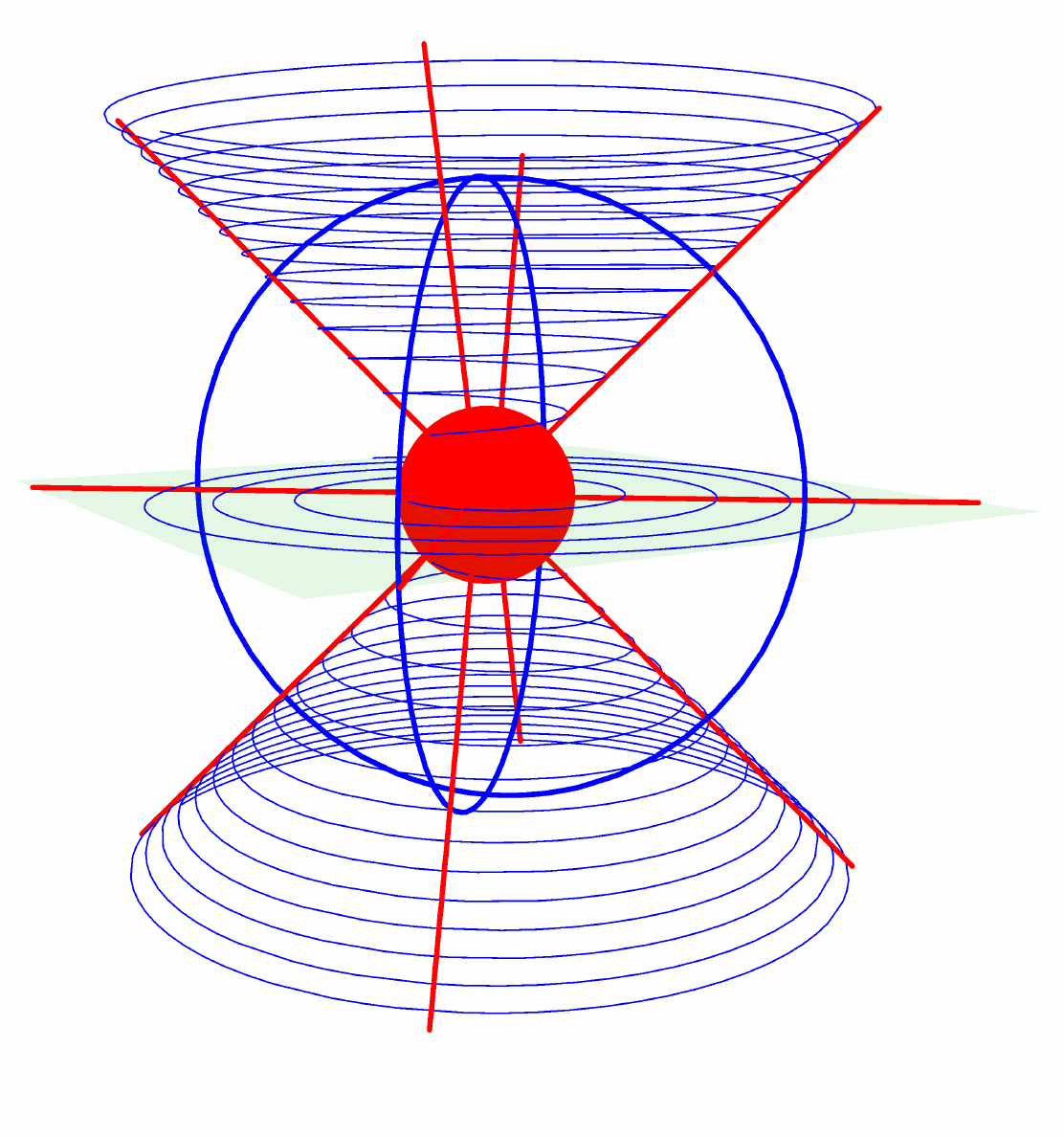} 
\put(5,10){\Large (a)}
\end{overpic}
\begin{overpic}[width=0.9\columnwidth]{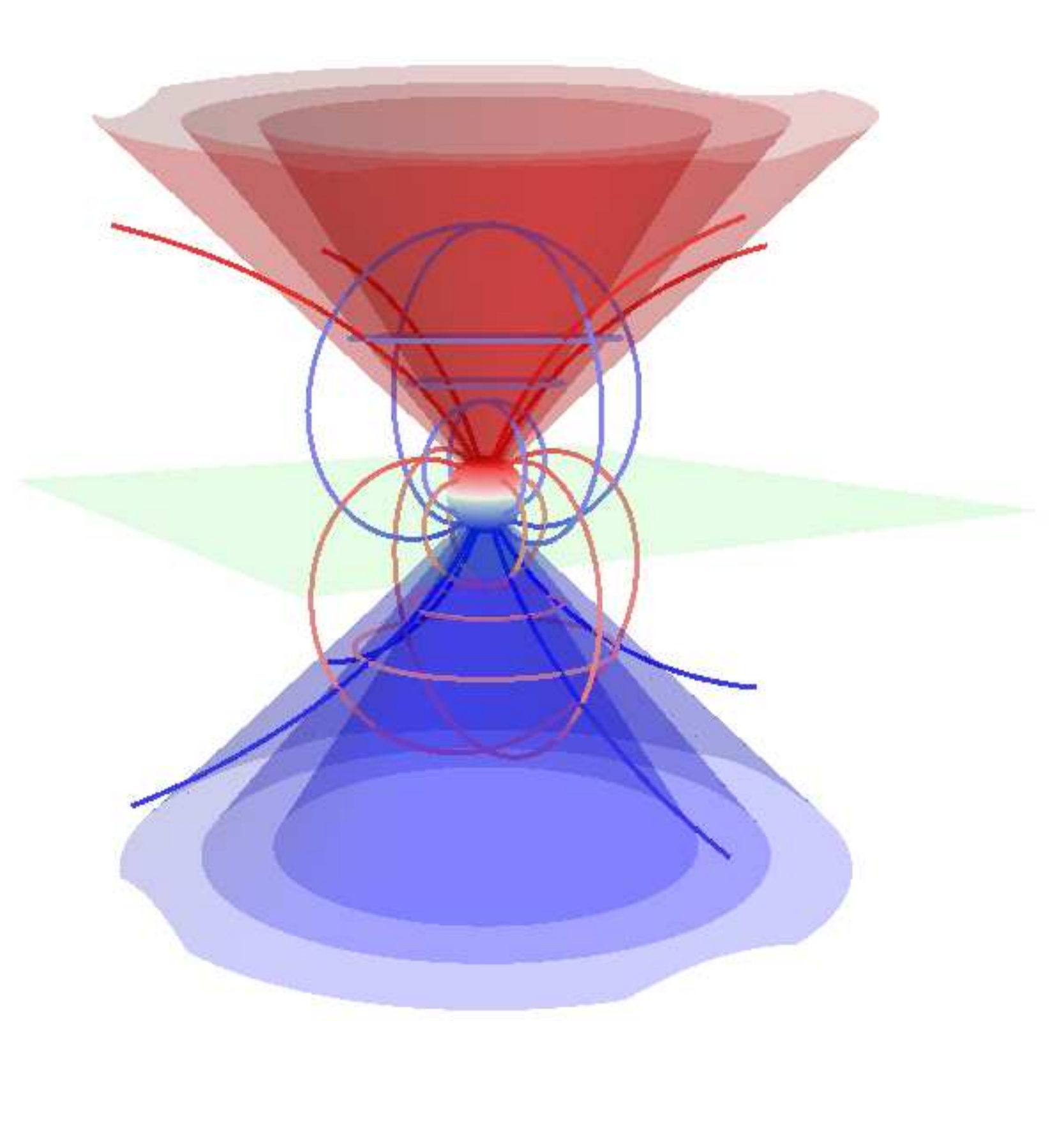}
\put(5,10){\Large (b)}
\end{overpic}
\caption{(a) Tendex lines, and (b) vortex lines 
for a slowly rotating (Kerr) black hole. Here we take $a/M = 0.1$.  The horizon is color coded by
its tendicity $\mathcal E_{NN}$ in (a) (uniformly red signifying negative tendicity) and vorticity $\mathcal B_{NN}$
in (b), and the field lines are color coded by the sign of their 
tendicity or vorticity (blue for positive, red for negative). That is, the radial tendex lines, the vortex lines emerging from the north pole and the azimuthal vortex lines on the bottom carry negative tendicity or vorticity, while all other lines have positive tendicity or vorticity.) In (a), the spiraling lines have been made to spiral more loosely by multiplying the rate of change in the $r$ direction by five. The semi-transparent cone-like surfaces emerging from the horizon's north and south polar regions show where the magnitude of the vorticity at a given radius has fallen to $80\%$ (outermost cones), $85\%$, and $90\%$ (innermost cones) of the polar magnitude. We identify the innermost cone (the 90\% contour) as the edge of the frame-drag vortex. The equatorial plane is shown for reference in both panels.}
\label{fig:SlowKerr}
\end{figure*}

The slow rotation gives rise to a frame-drag field
\begin{eqnarray}
\mathcal B_{\tilde r\tilde r} &=& \frac{-6aM\cos\theta}{r^4}\;, \quad
\mathcal B_{\tilde r\tilde\theta} = \mathcal B_{\tilde\theta\tilde r}
= \frac{-3aM\sin\theta}{r^4\sqrt{1+2M/r}}\;, \nonumber\\
\mathcal B_{\tilde\theta\tilde\theta} &=& \mathcal B_{\tilde\phi\tilde\phi}
=\frac{3aM\cos\theta}{r^4}\;
\label{Bijrot}
\end{eqnarray}
[Eq.~\eqref{eq:BLB} linearized in $a/M$] that lives in the slices of constant EF time $\tilde t$. 
This field's vortex lines, shown in Fig.\ \ref{fig:SlowKerr}b, 
are poloidal and closely 
resemble those of a spinning point mass (a ``current dipole'') 
in the linearized approximation
to general relativity (Fig.\ 3 of Paper I \cite{Nichols:2011pu}). 
At radii $r\gg M$, the field asymptotes to that of a 
linearized current dipole.

The rotating hole's horizon vorticity is $\mathcal B_{NN} = \mathcal B_{\hat r
\hat r} = -6(aM/r^4) \cos\theta$, which is negative in the north polar regions
and positive in the south polar regions.  Correspondingly, there is a 
counterclockwise frame-drag vortex sticking out of the hole's north pole,
and a clockwise one sticking out if its south pole. We 
identify the edge of each vortex, at radius $r$, as the location where the vorticities
of the vortex lines that emerge from the hole at the base of the vortex, 
fall (as a function of $\theta$ at fixed $r$) to $90\%$ of the on-pole vorticity. The vortex edges are shown, in Fig.\ \ref{fig:SlowKerr}, as semi-transparent surfaces; for comparison we also show where the vorticity has fallen to $85\%$ and $80\%$ of the on-pole vorticity at a given radius $r$. 

The hole's (small) spin not only generates a frame-drag field
$\mathcal B_{ij}$; it also modifies,
slightly, the hole's tidal field $\mathcal E_{ij}$ and its tendex lines.
However, the spin does not modify the field's tendicities, which 
(to first order in $a/M$) remain
$\lambda_r^{\mathcal E} = -2M/r^3$, $\lambda_\theta^{\mathcal E} 
= \lambda_\phi^{\mathcal E} = M/r^3$ [Eq.\ (\ref{eq:SchTendicities})]. 
The modified unit tangent vectors to the tendex lines are
\ba
\vec V^{\mathcal E}_r &=& \vec e_{\hat r} - {2Ma  \sin\theta \over
r^2 \sqrt{1+2M/r}} \vec e_{\hat{\tilde \phi}}\;, \nonumber\\
\vec V^{\mathcal E}_{\tilde \phi} &=& \vec e_{\hat {\tilde\phi}} + 
{2Ma  \sin\theta \over
r^2 \sqrt{1+2M/r}} \vec e_{\hat r}\;, \quad  
\vec V^{\mathcal E}_{\tilde\theta} = \vec e_{\hat{\tilde\theta}}\;.
\ea
Correspondingly, there is a slight (though hardly noticeable) bending of
the radial tendex lines near the black hole, and---more importantly---the
azimuthal tendex lines (the ones tangent to 
$\vec V^{\mathcal E}_{\tilde\phi}$) cease to close.  Instead, the azimuthal
tendex lines spiral outward along cones of fixed 
$\theta$, as shown in Fig.\ \ref{fig:SlowKerr}a. 
Since these lines have been only slightly perturbed from closed loops, they 
spiral quite tightly, appearing as solid cones. 
In order to better visualize these spiraling lines, we have increased their 
outward ($r$ directed) rate of change by a factor of five as compared to the 
axial rate of change in Fig.~\ref{fig:SlowKerr}a.

\subsection{Robustness of frame-drag field and tendex-line spiral}
\label{sec:SlowKerrRobustness}

The two new features induced by the hole's small spin (the frame-drag
field, and the spiraling of the azimuthal tendex lines) are, in fact,
robust under changes of slicing. We elucidate the robustness of the
tendex spiral in Appendix \ref{app:AzimuthalSpiral}.  
We here elucidate the robustness
of the frame-drag field and its vortex lines and vorticities:

Suppose that we change the time function $\tilde t$, which defines our time
slices, by a small fractional amount of order $a/M$; i.e., we introduce
a new time function
\be
t' = \tilde t + \xi( r,\theta)\;, 
\ee
where $\tilde t$ is EF time and $\xi$ has been chosen axisymmetric and 
time-independent, so it respects the symmetries of the black hole's spacetime. 
Then ``primed'' observers who move
orthogonal to slices of constant $t'$ will be seen by the EF observers
(who move orthogonal to slices of constant $\tilde t$) to have small 
3-velocities that are poloidal, ${\bf v} = v^{\hat r} {\bf e}_{\hat r}
+ v^{\hat \theta} {\bf e}_{\hat\theta}$.  The Lorentz transformation from
the EF reference frame to the primed reference frame at some event in 
spacetime induces a change of the frame-drag field given by  
\be
\delta \boldsymbol{\mathcal B} = -2({\bf v} \times \boldsymbol{\mathcal E})^S\;
\ee
[see, e.g., Eq.~(B12) of \cite{Maartens1998b}, linearized in small ${\bf v}$], 
where the $S$ means symmetrize.
Inserting the EF tidal field (\ref{eq:EFtidal}) and the poloidal components of 
$\bf v$, we obtain as the only nonzero components of $\delta {\mathcal B}$
\be
\delta \mathcal B_{\hat r \hat {\tilde \phi}} = 
\delta \mathcal B_{ \hat {\tilde \phi} \hat r} = -(3M/r^3) v^{\hat\theta}\;.
\ee

This axisymmetric, slicing-induced change of the frame-drag field does
not alter the nonzero components of the frame-drag field in Eq.\ 
(\ref{Bijrot}); it only introduces a change in the component 
$\mathcal B_{\hat r \hat {\tilde \phi}}$.
This is a sense in which we mean the frame-drag field is robust.
A simple calculation can show that one vorticity is unchanged,
$\mathcal B_{\hat\phi \hat\phi}=3aM\cos\theta/r^4$, but the corresponding 
vortex line will no longer be a circle of constant $(r,\theta)$.
Instead, it will wind on a sphere of constant $r$ relative to these closed 
azimuthal circles with an angle whose tangent is given by 
$v^{\hat \theta} \csc\theta \sqrt{r^2+2Mr}$.
The poloidal vortex lines must twist azimuthally to remain orthogonal to these
spiralling azimuthal lines, as well.

Although we will not see this specific kind of spiraling vortex lines
in the next section on rapidly rotating Kerr black holes, we will see a 
different spiraling of the azimuthal vortex lines: spiraling on cones of 
constant $\theta$.
We describe the reason for this in Appendix \ref{app:AzimuthalSpiral}.

\section{Rapidly Rotating (Kerr) Black Hole}
\label{sec:StaticKerr}

We shall now explore a rapidly rotating black hole described by the precise
Kerr metric.

\subsection{Kerr metric in Boyer-Lindquist coordinates}
\label{sec:KerrBL}

The Kerr metric is usually written in Boyer-Lindquist (BL) 
coordinates $\{t,r, \theta,\phi\}$, where it takes the form
\ba 
\label{eq:BLLineElem2}
ds^2 &=& -\left( 1-\frac{2Mr}{\Sigma} \right)dt^2 + \frac{\Sigma}{\Delta} dr^2 + \Sigma d\theta^2 + \frac{\sin^2\theta}{\Sigma} A d\phi^2
\notag\\
&& - \frac{4Mar\sin^2 \theta}{\Sigma} dtd\phi \,, \notag \\
\Sigma & = & r^2 + a^2 \cos^2 \theta \,, \notag \\
\Delta & =&  r^2 - 2 M r + a^2 \,,  \notag \\
A & =&  (r^2 + a^2)^2 - a^2 \Delta \sin^2 \theta\,,
\ea

Because the slices of constant $t$ are singular at the horizon (and
therefore not of much interest to us), we relegate to Appendix \ref{app:BLKerr}
the details of their tidal and frame-drag fields, and their vortex and tendex 
lines.

\subsection{Horizon-penetrating slices}

In our study of Kerr black holes, we shall employ two different slicings
that penetrate the horizon smoothly:
surfaces of constant Kerr-Schild time coordinate $\tilde t$, and surfaces
of constant Cook-Scheel time coordinate $\bar t$.  By comparing these two slicings' tendex lines with each other, and also their vortex lines with each other, 
we shall gain insight into the lines' slicing dependence.

The Kerr-Schild (\cite{Kerr1963,BoyerLindquist1967}, see also, e.g., Exercise 
33.8 of~\cite{MTW}) time coordinate (also sometimes called ingoing-Kerr time) 
is defined by
\be
\tilde t  = t + r_* -r\;, \quad
\hbox{where } 
\frac{dr_*}{dr}  = \frac{r^2 + a^2}{\Delta}\;.
\label{eq:tKS}
\ee
The Cook-Scheel~\cite{cook_scheel97} time coordinate is
\begin{align}
\bar{t} & = t + \frac{r_+^2+a^2}{r_+-r_-} \text{ln}\left|\frac{r-r_+}{r-r_-} 
\right|\; \nonumber \\
& = \tilde t + 2M \text{ln} \left| \frac{2M}{r-r_-} \right| \;,
\label{eq:tCS}
\end{align}
(see Eqs.\ (19) and (20) of \cite{cook_scheel97}) where $r_+$ is the value of the Boyer-Lindquist radial coordinate $r$ at the event horizon, and $r_-$ is its value at the (inner) Cauchy horizon:
\be
r_{\pm} = M \pm \sqrt{M^2-a^2}\;.
\label{eq:rpm}
\ee

\begin{figure}[t]
\vskip3pc
\includegraphics[width=0.9\columnwidth]{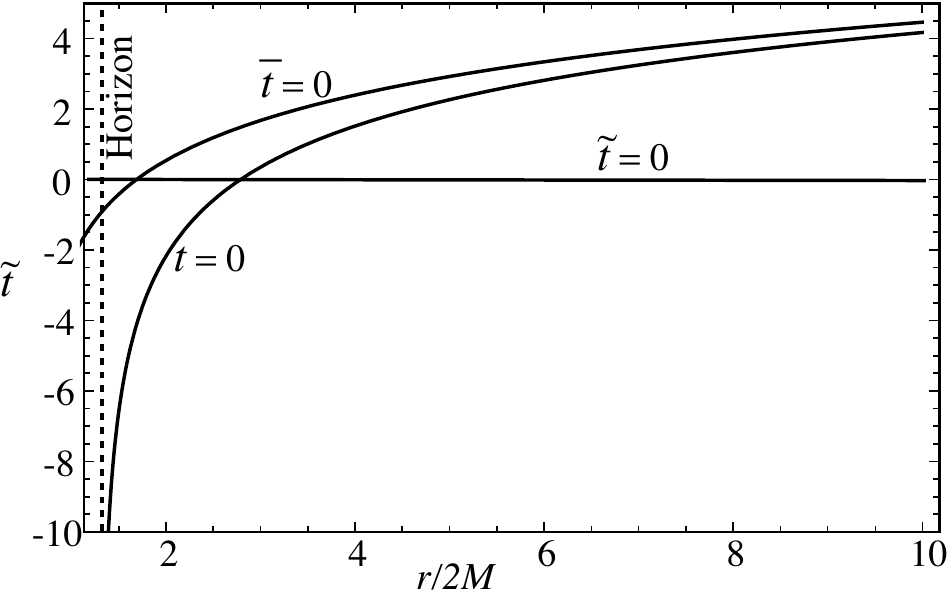}
\caption{Slices of constant Boyer-Lindquist time $t$, Kerr-Schild
time $\tilde t$, and Cook-Scheel time $\bar t$, drawn in a Kerr-Schild
spacetime diagram for a black hole with $a/M = 0.95$.
}
\label{fig:KerrSlicings}
\end{figure}

Figure \ref{fig:KerrSlicings} shows the relationship between these slicings for a black hole with $a/M= 0.95$.
In this figure, horizontal lines are surfaces of constant Kerr-Schild
time $\tilde t$.  Since $t$, $\bar t$ and $\tilde t$ differ solely by
functions of $r$, the surfaces of constant Cook-Scheel time $\bar t$ are all 
parallel to the $\bar t=0$ surface shown in the figure, and the surfaces
of constant Boyer-Lindquist time $t$ are all parallel to the $t=0$ surface.
The Kerr-Schild and Cook-Scheel surfaces penetrate the horizon smoothly.
By contrast, the Boyer-Lindquist surfaces all asymptote to the horizon 
in the deep physical past,
never crossing it; i.e., they become physically singular at the horizon.

\subsection{Horizon-penetrating coordinate systems}

Not only is the Boyer-Lindquist time coordinate $t$ singular at the event
horizon; so is the Boyer-Lindquist azimuthal angular coordinate $\phi$.  It
winds around an infinite number of times as it asymptotes to the horizon. 
We shall use two different ways to unwind it, associated with two different
horizon-penetrating angular coordinates:  The ingoing-Kerr coordinate
\be
\tilde \phi = \phi +  
\frac{a}{r_+-r_-} \text{ln}\left|\frac{r-r_+}{r-r_-} \right|
= \phi + \int_r^\infty {a\over\Delta} dr\;,
\label{eq:phiCS}
\ee
and the Kerr-Schild coordinate
\be
\varphi = \tilde \phi - \tan^{-1}(a/r)\;.
\label{eq:phiKS}
\ee

\begin{figure}[t]
\vskip3pc
\includegraphics[width=0.9\columnwidth]{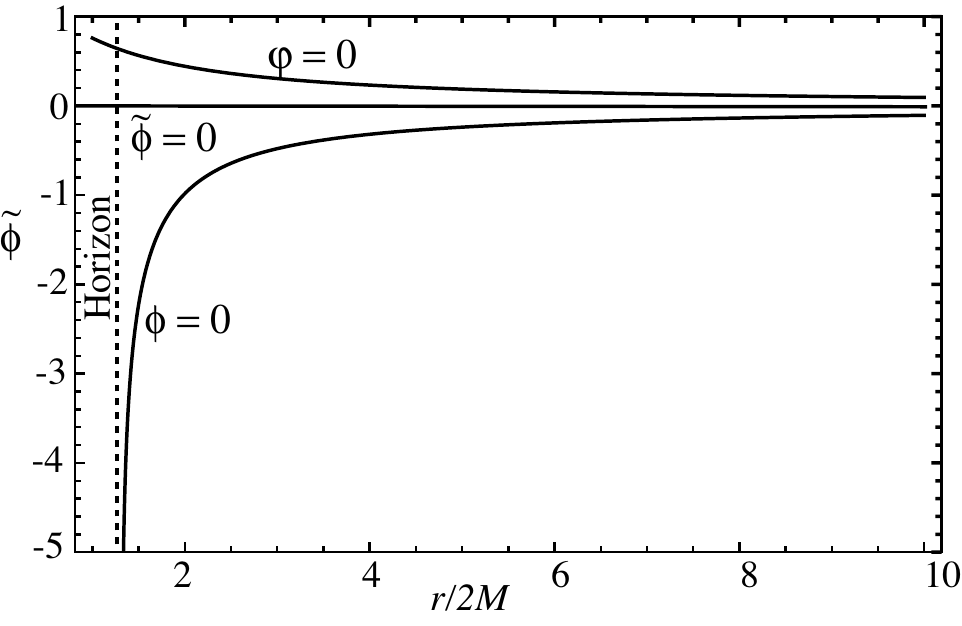}
\caption{Curves of constant Boyer-Lindquist angle $\phi$,
Kerr-Schild angle $\varphi$, and ingoing-Kerr angle $\tilde\phi$
for a black hole with $a/M = 0.95$.
}
\label{fig:KerrAngles}
\end{figure}

Figure \ref{fig:KerrAngles} shows the relationship of these angular
coordinates for a black hole with $a/m =0.95$. Notice that: (i) all three angular coordinates become
asymptotically the same as $r\rightarrow\infty$; (ii) 
the two horizon-penetrating coordinates, 
ingoing-Kerr $\tilde\phi$ and Kerr-Schild $\varphi$, differ by less than
a radian as one moves inward to the horizon; and (iii) the Boyer-Lindquist
coordinate $\phi$ plunges to $-\infty$ (relative to horizon-penetrating
coordinates) as one approaches the horizon, which means it wraps around
the horizon an infinite number of times. 

In the literature on Kerr black holes, four sets of spacetime coordinates
are often used:
\begin{itemize}

\item
\textit{\textbf{Boyer-Lindquist coordinates}}, 
$\{t,r,\theta,\phi\}$\;. These are the coordinates in Sec.\ \ref{sec:KerrBL}.

\item
\textit{\textbf{Ingoing-Kerr coordinates}}, 
$\{\tilde t, r,\theta,\tilde\phi\}$.  Often in this case $\tilde t$ is replaced 
by a null coordinate, $v=\tilde t + r$ (curves $v=$ const, $\theta=$ const, and 
$\tilde\phi=$ const are ingoing null geodesics).

\item
Quasi-Cartesian \textit{\textbf{Kerr-Schild coordinates}}, $\{\tilde t,
x,y,z\}$ and their cylindrical variant $\{\tilde t, \varpi, z, \varphi\}$.
Here 
\ba
x+iy &=& (r+ia)e^{i\tilde\phi}\sin\theta\;, \quad 
z = r \cos\theta\;, \notag \\
\varpi&=&\sqrt{x^2 + y^2} = \sqrt{r^2+a^2} \sin^2 \theta \;,  \notag \\
\varphi &=& \arctan(y/x) = \tilde\phi + \arctan(a/r)\;.
\label{eq:KSxy}
\ea
The Kerr-Schild spatial coordinates $\{x,y,z\}$ 
resemble the coordinates typically used in numerical simulations
of binary black holes at late times, when the merged hole is settling down
into its final, Kerr state. These coordinate systems 
resemble each other in the senses that 
(i) both are quasi-Cartesian, and (ii)
for a fast-spinning hole, the event horizon in both cases,
when plotted in the coordinates being used, looks moderately oblate.
For this reason, in our study of Kerr black holes,
we shall focus our greatest attention on Kerr-Schild coordinates.
The Kerr metric, written in Kerr-Schild coordinates, has the form
\ba
ds^2 & = & \left(\eta_{\mu \nu} +  \frac{2Mr^3}{r^4+a^2 z^2} k_\mu k_\nu \right) dx^\mu dx^\nu \,, \notag \\
k_\mu & =& \left( 1, \frac{r x + a y}{r^2 + a^2}, \frac{r y - a x}{r^2 + a^2}, \frac{z}{r} \right) \,,
\ea
where $r$ is the Boyer-Lindquist radial coordinate, and is 
the larger root of
\ba
x^2 + y^2 + z^2 = r^2 + a^2\left( 1-\frac{z^2}{r^2} \right)\,,
\ea
and $\eta_{\mu \nu}$ is the usual flat Minkowski metric.

\item
\textit{\textbf{Cook-Scheel harmonic coordinates}} \cite{cook_scheel97},
$\{\bar t, \bar x, \bar y, \bar z\}$ where $\bar t$ is given by Eq.~\eqref{eq:tCS}, 
while the spatial coordinates are defined by 
\ba
\bar{x} + i\bar{y} &=& \left[ r -M + ia \right] e^{i \tilde{\phi}} \sin(\theta) \\
\bar{z} &=& \left[ r -M \right] \cos(\theta)
\ea
These coordinates are harmonic in the sense that
 the scalar wave operator acting on them vanishes.  
In these coordinates, the event horizon of a spinning black hole
is more oblate than in Kerr-Schild coordinates---and much more oblate for 
$a/M$ near unity. 
\end{itemize}

\subsection{Computation of tendex and vortex lines, and their tendicities
and vorticities}

Below we show pictures of tendex and vortex lines, color coded with
their tendicities and vorticities, for our two horizon-penetrating slicings
and using our three different sets of spatial coordinates.  In all cases
we have computed the field lines and their eigenvalues numerically, 
beginning with analytical formulas for the metric. More specifically, 
after populating a numerical grid using analytical expressions for the metric, 
we numerically compute $\mathcal E_{ij}$ and
$\mathcal B_{ij}$, as well as their eigenvalues and eigenvectors. 
A numerical integrator is then utilized to generate the tendex and vortex lines.
Finally, we apply analytical transformations that take these lines to whatever
spatial coordinate system we desire.  

Although not required for the purpose of generating the figures in the following sections, 
it is nevertheless possible to find analytical expressions for $\mathcal E_{ij}$ and
$\mathcal B_{ij}$, and subsequently their eigenvalues and eigenvectors. 
These expressions provide valuable insights into the behavior of the tendex and vortex lines, and we present such results for the ingoing-Kerr coordinates 
in Appendix \ref{app:IngoingKerr}.

\subsection{Kerr-Schild slicing: Tendex and vortex lines in several
spatial coordinate systems}

Once the slicing is chosen, the tidal and frame-drag fields, and also the
tendex and vortex lines and their tendicities and vorticities, are all
fixed as geometric, coordinate-independent entities that live in a slice.  
If we could draw
an embedding diagram showing the three-dimensional slice isometrically embedded
in a higher-dimensional flat space, then we could visualize the tendex 
and vortex lines without the aid of a coordinate system.  However, the
human mind cannot comprehend embedding diagrams in such high-dimensional
spaces, so we are forced to draw the tendex and vortex lines in some
coordinate system for the slice, in a manner that makes the coordinate
system look like it is one for flat space.  

Such a coordinate-diagram plot of the lines makes them look coordinate
dependent---i.e., their shapes depend on the coordinate system used.
Nevertheless, the lines themselves are geometrically well-defined, independent 
of coordinate system, and they map appropriately between them. The 
visual features of these lines are also qualitatively similar in 
reasonable coordinate systems.

\begin{figure*}[tbp]
\includegraphics[width=0.9\textwidth]{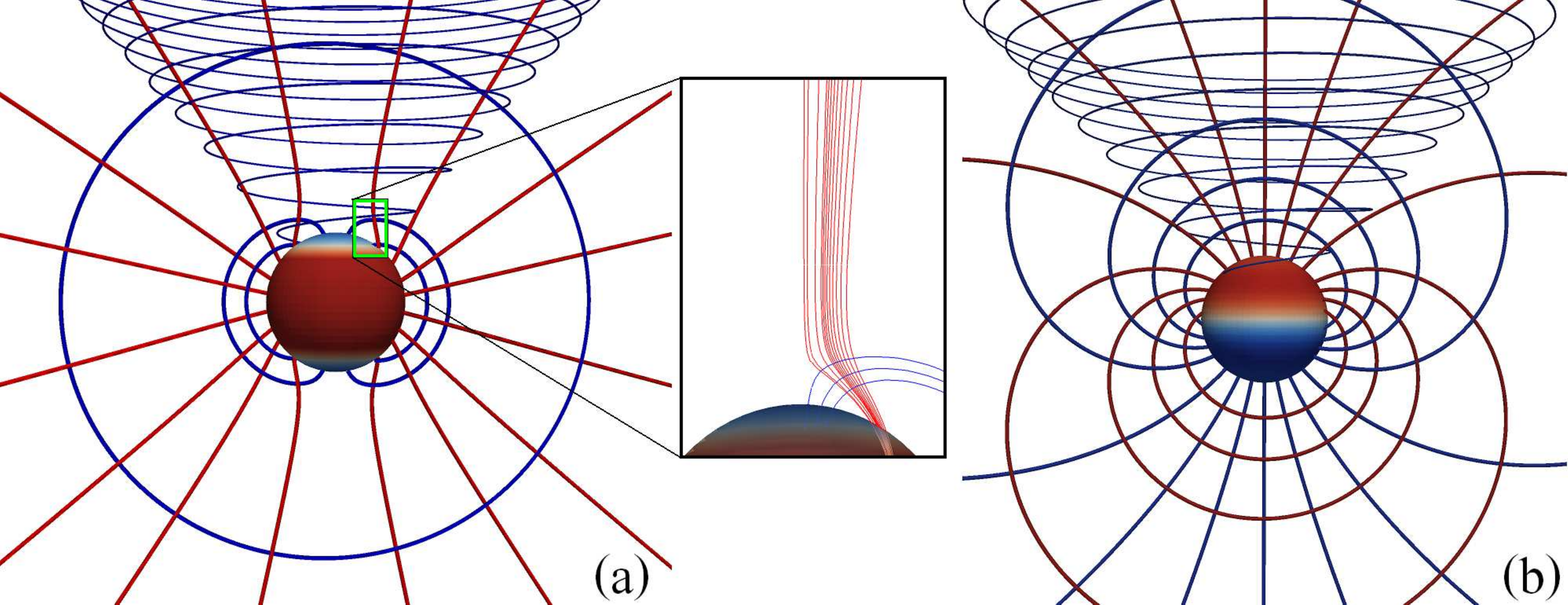}
\includegraphics[width=0.9\textwidth]{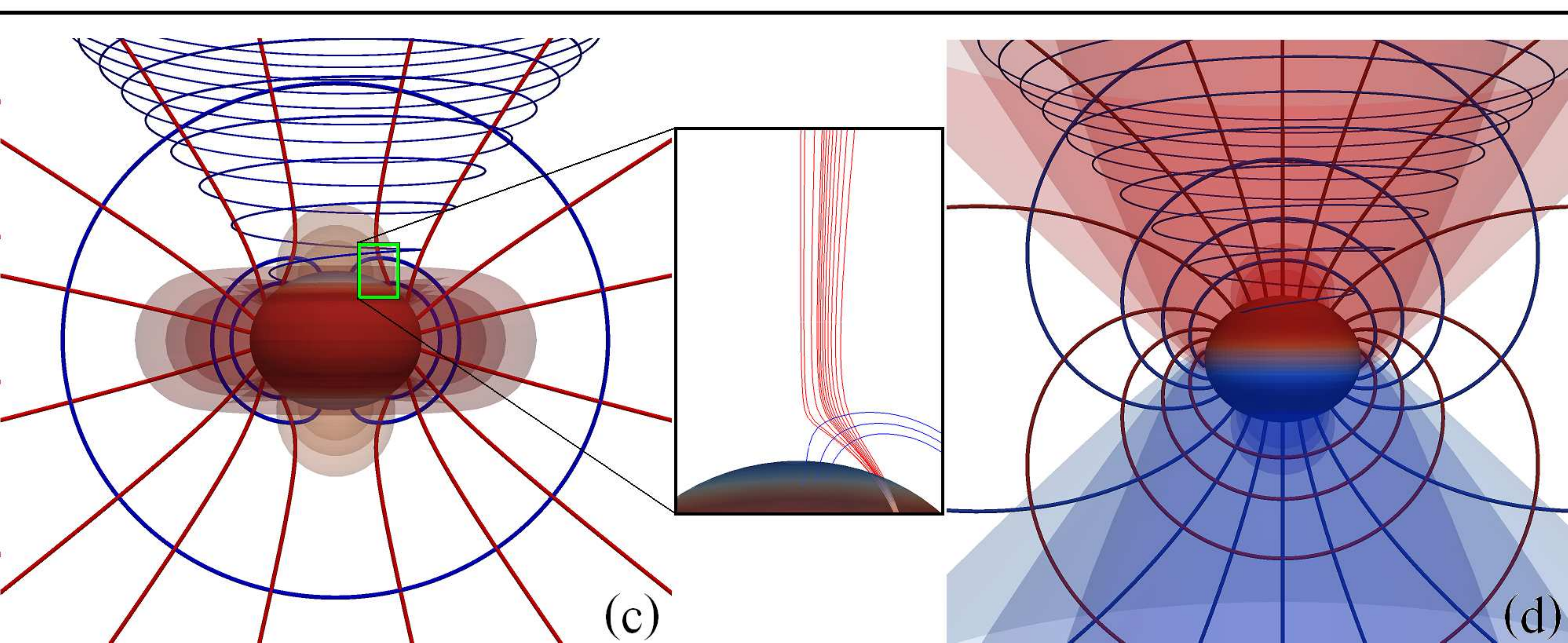}
\includegraphics[width=0.9\textwidth]{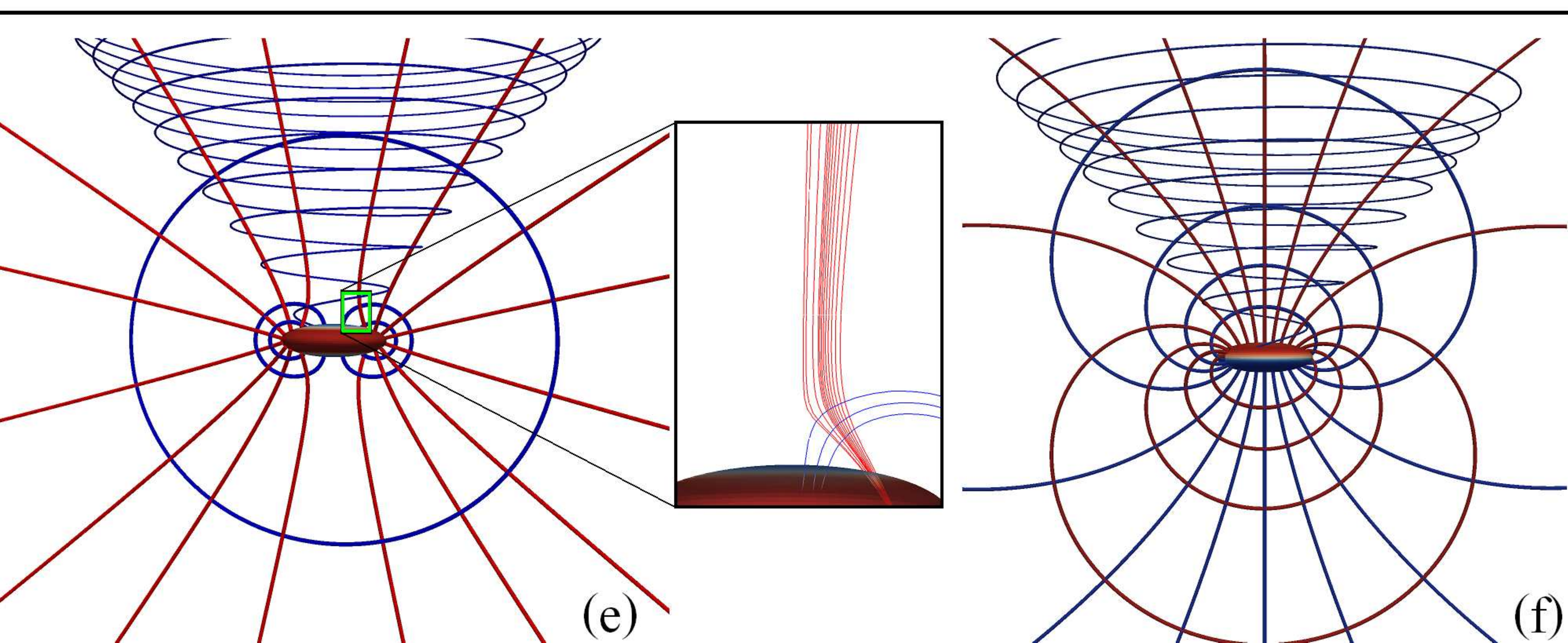}
\caption{Kerr  
black hole with $a/M=0.95$ in \textit{\textbf{Kerr-Schild slicing}},
drawn in three different spatial coordinate systems.  The left and center
columns of drawings [panels (a), (c), (e)] show tendex lines; the right
column of drawings [panels (b), (d), (f)] show vortex lines.  The three
rows, from top downward, use ingoing-Kerr spatial coordinates $\{r,\theta,
\tilde \phi\}$ [panels (a) and (b)], Kerr-Schild spatial coordinates 
$\{x,y,z\}$ [panels (c) and (d)], and Cook-Scheel spatial coordinates
$\{\bar x, \bar y, \bar z\}$ [panels (e) and (f)]. 
In all cases, the lines with positive
tendicity or vorticity are colored blue; those with negative tendicity or
vorticity are colored red. (The radial tendex lines and vortex lines emerging from the top half of the horizon are red, while all other lines are blue.) The horizon is shown with its horizon
tendicity (left column of drawings) and horizon vorticity (right column)
color coded from dark blue for strongly positive to dark red for strongly
negative. (Horizon tendicity is negative near the equator and positive near the poles. Vorticity on the other hand transitions from being negative on top to positive on the bottom.) 
In Kerr-Schild coordinates [panels (c) and (d)], 
we have also shown as semi-transparent surfaces,  
contours of $\tilde{r}^3$ times tendicity and $\tilde{r}^4$ times vorticity, where $\tilde r^2 = x^2+y^2+z^2$ for Kerr-Schild spatial coordinates. In panel (c), the innermost equatorial contour has the most negative tendicity while the others have $90\%,80\%, 30\%, 20\%,$ and $10\%$ this value, and the innermost polar contour has the least negative tendicity. In panel (d) the contour with the most negative vorticity consists of the innermost red cone and the outermost red bubble (at the north pole), and the others are at $90\%$ and $80\%$ this value. The blue contours of panel (d) (at the bottom half of that panel) are arranged similarly but with positive vorticity. 
}
\label{fig:CompareCoords}
\end{figure*}

Figure \ref{fig:CompareCoords} is an important example.  
It shows the tendex lines (left
column of plots) and vortex lines (right column of plots) for a fast-spinning
Kerr black hole, with $a = 0.95 M$. We have also colored the horizon of the 
black hole according to its horizon tendicity and vorticity, respectively.  
In all cases the slicing is 
Kerr-Schild; i.e., the lines lie in a slice of constant $\tilde t$.
The three rows of figures are drawn in three different spatial coordinate
systems: ingoing-Kerr, Kerr-Schild, and Cook-Scheel.  

Notice the following important features of this figure:
\begin{itemize}
\item
As expected, the qualitative features of the
tendex lines are independent of the spatial coordinates.
The only noticeable differences from one coordinate system to another
are a flattening of the strong-gravity region near the hole as one
goes from ingoing-Kerr coordinates (upper row of panels) to 
Kerr-Schild coordinates (center row of panels) and then a further flattening
for Cook-Scheel coordinates (bottom row of panels).  
\item
The azimuthal (toroidal) tendex and vortex lines (those that point 
predominantly in the $\vec e_{\tilde \phi}$ direction) spiral outward from the
horizon along cones of constant $\theta$, as for the tendex
lines of a slowly spinning black hole 
[cf.\ the form of $\vec V_{\hat{\tilde\phi}}$
in Eqs.\ (\ref{KSEigenVecs})]. As we shall discuss in Appendix
\ref{app:AzimuthalSpiral},
this is a characteristic of a large class of commonly used, horizon-penetrating
slicings of spinning black holes.
\item
All the poloidal tendex and vortex lines have (small) azimuthal ($\tilde \phi$)
components, which do not show up in this figure; see the $\vec e_{\hat{\tilde
\phi}}$ components of the eigenvectors $\vec V^{\mathcal E}_r$,
$\vec V^{\mathcal E}_\theta$, $\vec V^{\mathcal B}_-$ and 
$\vec V^{\mathcal B}_+$ in Eqs.\ (\ref{KSEigenVecs}) and 
(\ref{KSBEigenVecs}).  
\item
Left column of drawings: 
For this rapidly spinning black hole, the horizon tendicity is positive
(blue) in the north and south polar regions, and negative (red) 
in the equatorial region, by contrast with a slowly spinning hole, where
the horizon tendicity is everywhere negative (Fig.\ \ref{fig:SlowKerr}). 
Correspondingly, a radially oriented person falling into a polar region
of a fast-spinning
hole gets \emph{squeezed} from head to foot, rather than stretched, as
conventional wisdom demands. The relationship $\mathcal E_{NN} = 
- \mathcal R/2$ between the horizon's tendicity and its scalar curvature
tells us that this peculiar polar feature results from the well-known
fact that,
when 
the spin exceeds $a/M = \sqrt{3}/2 \approx 0.8660$,  
the scalar curvature goes negative near the poles, at angles
$\theta$ satisfying 
$2 (a/M)^2 \cos^2 \theta > 1 + \sqrt{1 - (a/M)^2}$. 
This negative scalar curvature is also
responsible for the fact that is it impossible to embed the horizon's 2-geometry in
a 3-dimensional Euclidean space when the spin exceeds $a/M = \sqrt{3}/2$~\cite{Smarr:1973}. 
\item
Left column of drawings: 
The blue (positive tendicity) tendex lines that emerge from the north
polar region sweep around the hole, just above the horizon, and descend
into the south polar region.  In order to stay orthogonal to these blue
(squeezing) tendex lines, the red (stretching) lines descending from radial
infinity get deflected away from the horizon's polar region until they 
reach a location with negative tendicity (positive scalar curvature), 
where they can attach to the horizon; see the central panels, which are
enlargements of the north polar region for the left panels. 
\item
Right column of drawings:
The vortex-line structure for this fast-spinning black hole is very
similar to that for the slow-spinning hole of Fig.\ 
\ref{fig:SlowKerr}, and similar to that for a spinning point mass in
the linear approximation to general relativity (Fig.\ 3 of Paper I \cite{Nichols:2011pu}).
The principal obvious change is that the azimuthal vortex lines are
not closed; instead, they spiral away from the black hole, like
the azimuthal tendex lines.  
\item
Right column of drawings:
Most importantly, as for a slow-spinning black hole, there are two
vortexes (regions of strong vorticity): as a counterclockwise vortex
emerging from the north polar region, and a clockwise vortex emerging
from the south polar region.  
As we shall see in Paper IV, when
two spinning black holes collide and merge, these vortexes sweep
around, emitting gravitational waves.
In Fig.~\ref{fig:CompareCoords}(d), these vortexes are indicated by 
contours of $\tilde{r}^4$ times vorticity, where $\tilde{r}^2=x^2+y^2+z^2$ for Kerr-Schild coordinates  $\{x,y,z\}$.
Notice in particular that each contour consists of one cone together with 
one bubble attached to the horizon, with the bubbles 
enclosing the polar regions excluding them from the vortexes. 
This is a feature not seen for the slow-spinning case.  

\end{itemize}

\subsection{Slicing-dependence of tendex and vortex lines}
\label{sec:KerrSlicingDependence}

\begin{figure*}[tbp]
\includegraphics[width=0.9\textwidth]{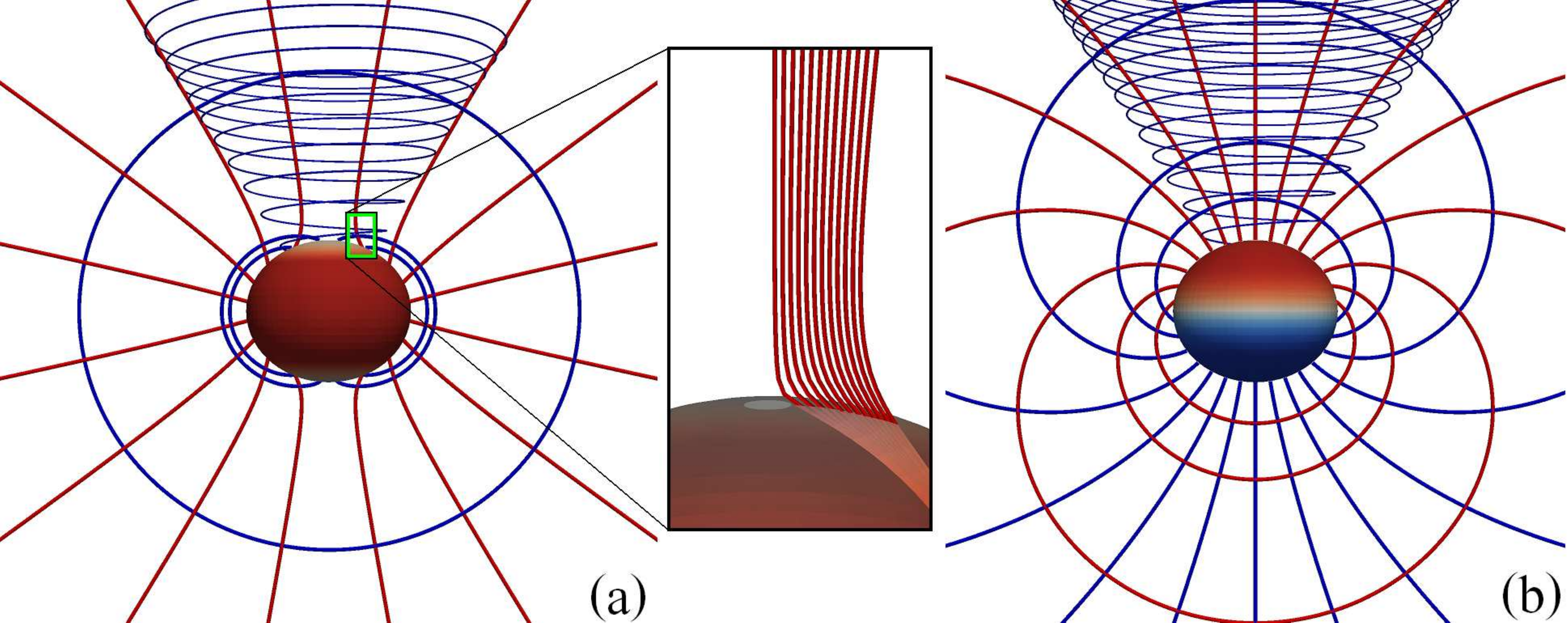}
\includegraphics[width=0.9\textwidth]{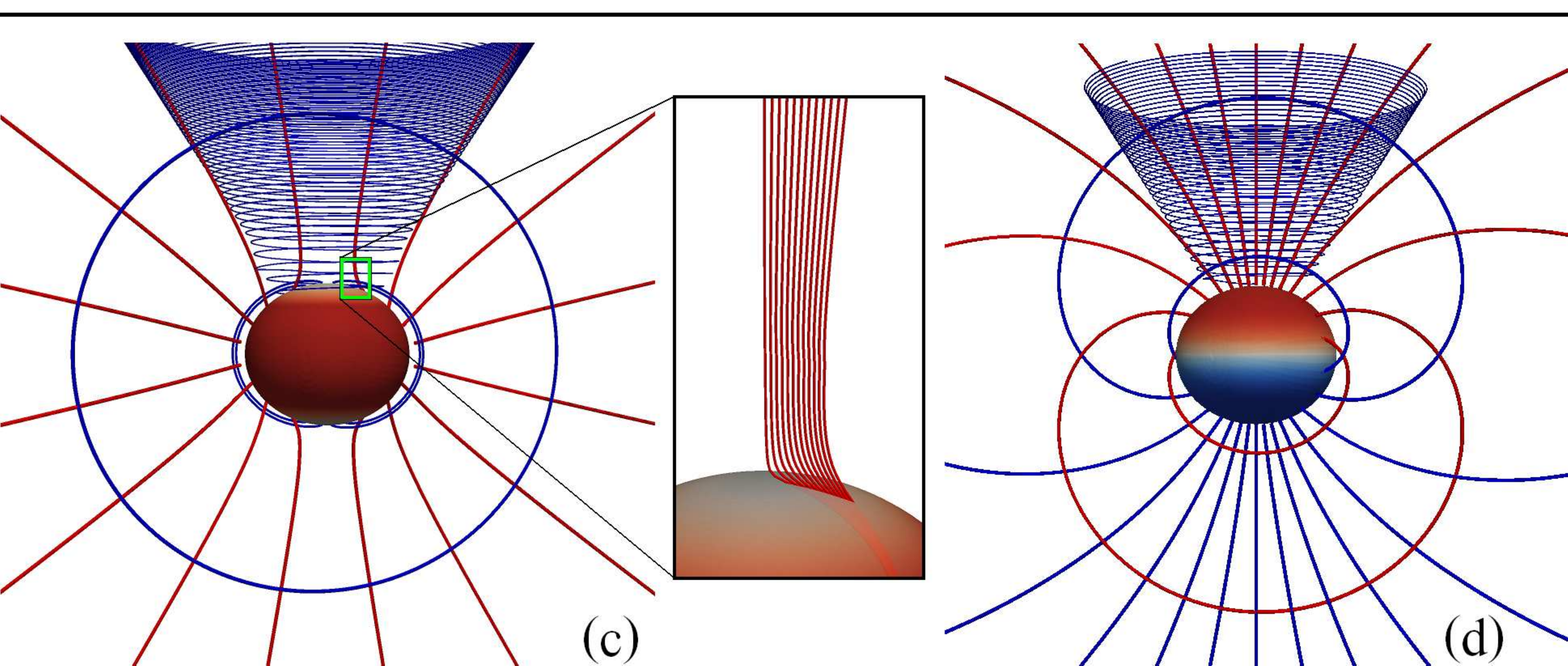}
\caption{
Tendex lines and vortex lines for a Kerr
black hole with $a/M=0.875$ in \textit{\textbf{Kerr-Schild spatial coordinates}},
for two different slicings: \textit{\textbf{Kerr-Schild}} $\tilde t =$constant,
and \textit{\textbf{Cook-Scheel }}$\bar t =$constant. 
The left and center
columns of drawings [panels (a) and (c)] show tendex lines; the right
column of drawings [panels (b) and (d)] show vortex lines.  The top row
of drawings [panels (a) and (b)] is for Kerr-Schild slicing; the bottom
row [panels (c) and (d)] is for Cook-Scheel slicing. Since the slicings
are different, it is not possible to focus on the same sets of field lines
in the Kerr-Schild (upper panels) and Cook-Scheel (lower panels) cases.
However,
we have attempted to identify similar field lines by ensuring they pass
through the same Kerr-Schild spatial coordinate locations on selected
surfaces. (The color of the lines and horizon are similar to Fig. 5)}
\label{fig:CompareSlicing}
\end{figure*}

To explore how a Kerr black hole's vortex and tendex lines depend on the
choice of slicing, we focus in Fig.\ \ref{fig:CompareSlicing} on a black hole 
with $a/M = 0.875$, viewed in a slice of constant Kerr-Schild time, 
$\tilde t =$constant, and in a slice of constant Cook-Scheel harmonic time, 
$\bar t =$constant. 
In the two slices, we use the same spatial coordinates: Kerr-Schild.  
(We chose $a/M =0.875$, rather than
the $0.95$ that we used for exploring spatial coordinate dependence, 
because it is simpler to handle numerically in the Cook-Scheel slicing.) 

The most striking aspect of Fig.\ \ref{fig:CompareSlicing} is the
close similarity of the tendex lines (left column of drawings) in the
two slicings (upper and lower drawings), and also the close similarity
of the vortex lines (right column of drawings) in the two slicings
(upper and lower). There appears to be very little slicing dependence 
when we restrict ourselves to horizon-penetrating slicings.  

By contrast,
if we switch from a horizon-penetrating to a horizon-avoiding
slice, there are noticeable changes in the field lines:  
Compare the top row of Fig.~\ref{fig:CompareSlicing}
($a/M =0.875$ for a Kerr-Schild, horizon-penetrating slice) with Fig.\
\ref{fig:BL} (the same hole, $a/M =0.875$, for a Boyer-Lindquist, 
horizon-avoiding slice), concentrating for now on panels (a) and (b) depicting 
tendex and vortex lines in Boyer-Lindquist spatial coordinates.
The most striking differences are 
(i) the radial tendex lines' failure to reach the horizon 
for horizon-avoiding slices,
contrasted with their plunging through the horizon for horizon-penetrating
slices, and (ii) the closed azimuthal tendex and vortex lines
for Boyer-Lindquist horizon-avoiding slices, contrasted with the
outward spiraling azimuthal lines for horizon-penetrating slices.  
In Appendix \ref{app:AzimuthalSpiral} we show that this outward spiral is 
common to a class of horizon-penetrating slices.
Lastly, we note that Fig.~\ref{fig:BL} (a) and (b) are plotted using 
Boyer-Lindquist spatial coordinates in order to compare
with analytical expressions given in that appendix. When we use Kerr-Schild spatial coordinates, as is done in Fig.~\ref{fig:CompareSlicing}, in order to facilitate a more appropriate 
comparison, we observe that the Boyer-Lindquist 
azimuthal coordinate singularity depicted in Fig.~\ref{fig:KerrAngles} causes 
the tendex and vortex lines in Boyer-Lindquist slicing to wind in $\phi$ direction when close to horizon. This feature is clearly visible in Fig.~\ref{fig:BL} (c) and (d), where we display the tendex and vortex
lines in Kerr-Schild spatial coordinates.

Based on our comparison of Kerr-Schild and Cook-Scheel slicings 
(Fig.\ \ref{fig:CompareSlicing}), and our analysis of the ubiquity
of azimuthal spiraling lines in horizon-penetrating slices (Appendix
\ref{app:AzimuthalSpiral}),
we conjecture that horizon-penetrating slicings of any black-hole spacetime 
will generically share the same qualitative and semi-quantitative structures of 
tendex and vortex lines. This conjecture is of key importance for 
our use of tendex and vortex lines to 
extract intuition into the dynamical processes observed in numerical 
simulations.  More specifically:

Numerical spacetimes have dynamically chosen slicings, and the primary 
commonality from simulation to simulation is that the time slicing 
must be horizon penetrating, to prevent coordinate singularities 
from arising on the numerical grid near the horizon. Our conjecture implies 
that, 
regardless of the precise slicing used in a simulation, we expect the 
tendex and vortex lines to faithfully reveal the underlying physical 
processes. We will build more support for this conjecture in
Paper III,
by comparing the final stages of a numerical black-hole merger 
with a perturbed Kerr black hole, using very different slicing 
prescriptions. 

We conclude this section with a digression from its slicing-dependence
focus: 

When we compare the $a/M =0.875$ black hole of Fig.\ \ref{fig:CompareSlicing}
with the $a/M = 0.95$ hole of Fig.\ \ref{fig:CompareCoords}, the
most striking difference is in the tendex lines very near the horizon.
The value 
$a/M = 0.875$ is only slightly above the critical spin 
$a/M = \sqrt{3}/2 = 0.8660$ at which the horizon's poles acquire negative
scalar curvature.  Correspondingly, for $a/M =0.875$, the blue tendex lines
that connect the two poles emerge from a smaller region at the poles
than for $a/M = 0.95$, and they hug the horizon more tightly as they
travel from one pole to the other; and the red, radial tendex lines 
near the poles suffer much smaller deflections than for $a/M = 0.95$
as they descend into the horizon (see insets).

\section{Conclusion}
\label{Conclusion}

Using vortex and tendex lines and their vorticities and tendicities, we have
visualized the spacetime curvature of stationary black holes. 
Stationary black-hole spacetimes are a simple arena in which to learn about the
properties of these visualization tools in regions of strong spacetime 
curvature.
From the features of the vortex and tendex lines and their vorticities and
tendicities that we describe below, we have gained an understanding of these 
visualization tools and made an important stride toward our larger goal of 
using these tools to identify geometrodynamical properties of strongly curved 
spacetimes---particularly those in the merger of binary black holes.

Black hole spacetimes have an event horizon (a feature that was absent in our
study of weakly gravitating systems in Paper I).
To understand our visualization tools on the horizon, we defined and discussed 
the {\it horizon tendicity} and {\it horizon vorticity} of stationary black
holes.
The horizon tendicity and vorticity are directly proportional to the intrinsic
and extrinsic curvature scalars of a two-dimensional horizon.
As a result, the average value of the horizon tendicity must be negative, and
the horizon vorticity must average to zero.
Any region of large vorticity on the horizon (a horizon vortex), therefore,
must be accompanied by an equivalent vortex of the opposite sign, but there 
is not an analogous constraint for horizon tendexes.

Outside the horizon, we also visualized the tendex lines and vortex lines, 
the tendicities and vorticities, and the regions of large tendicity 
({\it tendexes}) and large vorticity ({\it vortexes}) for Schwarzschild and
Kerr black holes (the latter both slowly and rapidly spinning).
In particular, we investigated how the vortex and tendex lines of 
Kerr black holes changed when they were drawn in different time slices and 
with different spatial coordinates---within the set of those time slices that 
smoothly pass through the horizon and spatial coordinates that are everywhere 
regular.
We found our visualizations are quite similar between two commonly used, 
though rather different, horizon-penetrating time functions: Kerr-Schild and 
Cook-Scheel.
The spatial-coordinate dependence was also mild, and was easily understandable
in terms of the relation between the different coordinate systems.
Because the coordinate systems used in numerical simulations of black holes 
are also horizon penetrating, this suggests that the vortex and tendex lines 
will not be very different, even though the dynamical coordinates of the 
simulation may be.

This study is a foundation for future work on computing the tendexes and 
vortexes of black-hole spacetimes. 
A recent work by Dennison and Baumgarte~\cite{Dennison2012}---in which the 
authors calculated the tendex and vortex fields of approximate initial data 
representing non-spinning, boosted black holes, and also black-hole 
binaries---will also be helpful for understanding binaries. In addition, our investigations of the slicing and coordinate dependence of tendexes and vortexes is complemented by another recent study of Dennison and Baumgarte \cite{Dennison:2012vf}, where expressions are given for computing curvature invariants in terms of the vorticities, tendicities, and the eigenvector fields which give the tendex and vortex lines. These expressions will likely be of use in future analytic and numerical studies of tendexes and vortexes.

In a companion paper (Paper III), we turn to perturbed black holes. 
We aim to deepen our understanding of tendex and vortex lines in these 
well-understood situations and to see what new insights we can draw from 
these spacetimes by using vortex and tendex lines. 
Ultimately, we will apply these visualization techniques and our intuition 
from simpler analytical spacetimes to study numerical simulations of strongly 
curved and dynamic spacetimes and their geometrodynamics.
In Paper IV, we will do just this, focusing on binary-black-hole mergers.

\acknowledgments

We thank Jeandrew Brink and Jeff Kaplan for helpful discussions. We would like to thank Mark Scheel for helpful discussions and for version control assistance, and B\'{e}la Szil\'{a}gyi for his help on numerical grid construction. A.Z. would also like to thank the National Institute for Theoretical Physics of South Africa for hosting him during a portion of this work.
Some calculations have been performed using the Spectral Einstein Code ({\tt SpEC})~\cite{SpECwebsite} on the Caltech computer cluster \textsc{zwicky}. This research was supported by NSF grants PHY-0960291, PHY-1068881 and CAREER grant PHY-0956189 at Caltech, by NSF grants PHY-0969111 and PHY-1005426 at Cornell, by NASA grant NNX09AF97G at Caltech, by NASA grant NNX09AF96G at Cornell, and by the Sherman Fairchild Foundation at Caltech and Cornell, the Brinson Foundation at Caltech, and the David and Barbara Groce fund at Caltech. 

\appendix

\section{Kerr Black Hole in Boyer-Lindquist Slicing and Coordinates} 
\label{app:BLKerr}

For a rapidly rotating Kerr black hole in Boyer-Lindquist (BL) coordinates
$\{t,r,\theta,\phi\}$, the metric is given by Eq.\ (\ref{eq:BLLineElem2}) 
above.  
A ``BL observer'', who moves orthogonally to the slices of constant BL time $t$,
has a 4-velocity $\vec u$ and orthonormal tetrad given by
\begin{align}
\label{BLCoordTetrad}
\vec{u} & = \sqrt{\frac{A} {\Sigma \Delta}}\left( \partial_t - \frac{2Mar}{A} \partial_\phi \right)  \,, &\vec{e}_{\hat r} & =  \sqrt{\frac{\Delta} {\Sigma}} \partial_r \,, \notag \\
\vec{e}_{\hat \theta} & =  \frac{1}{\sqrt{\Sigma}} \partial_\theta \,, &
\vec{e}_{\hat \phi} & = \sqrt{\frac{\Sigma}{A}}\frac{1}{\sin \theta} \partial_\phi \, .
\end{align}
This tetrad is also often called the locally nonrotating frame 
\cite{Bardeen1970, 1972ApJ...178..347B}.
In this orthonormal basis, the tidal and frame-drag fields are given
by (cf.\ Eqs.\ (6.8a-6.9d) of \cite{Thorne-Price-MacDonald:Kipversion})
\begin{subequations}
\ba
\mathcal{E}_{\hat a \hat b}&  =& \bma -Q_e \frac{2+\xi}{1-\xi} & \mu Q_m & 0 \\ * & Q_e \frac{1+2\xi}{1-\xi} & 0 \\ * & * & Q_e \ema \,, \label{eq:BLE}
\\
\mathcal{B}_{\hat a \hat b} & =& \bma -Q_m \frac{2+\xi}{1-\xi} & -\mu Q_e & 0 \\ * & Q_m \frac{1+2\xi}{1-\xi} & 0 \\ * & * & Q_m \ema \,, \label{eq:BLB}
\ea
with entries denoted by $*$ fixed by the symmetry of the tensors, and where
\ba
&&Q_e = \frac{Mr(r^2-3a^2 \cos^2\theta)}{\Sigma^3}  \,,
\label{eq:Qe}
\\
&&Q_m = \frac{Ma\cos\theta(3r^2-a^2 \cos^2\theta)}{\Sigma^3} \,,
\label{eq:Qm}
\\
&& \xi = \frac{\Delta a^2 \sin^2 \theta}{(r^2+a^2)^2} \,, 
\label{eq:Xi} \\
&& \mu = \frac{3a\sqrt{\Delta} (r^2+a^2)\sin \theta}{A} 
= \frac{3\sqrt \xi}{1-\xi} \,. 
\label{eq:MuXi}
\ea
\end{subequations}
The functions $Q_e$ and $Q_m$ are related to the real and imaginary parts of 
the complex Weyl scalar $\Psi_2$ calculated using the Kinnersley null tetrad 
by $\Psi_2 = -Q_e + i Q_m$. 
Note that there is a duality between the electric and the magnetic curvature 
tensors: namely, by replacing $Q_e \to Q_m$ and $Q_m \to -Q_e$, the tensor
transforms as $\mathcal{E}_{\hat a \hat b} \to \mathcal{B}_{\hat a \hat b}$. 

The block diagonal forms of $\mathcal{E}_{\hat a \hat b}$ and 
$\mathcal B_{\hat a \hat b}$ imply that one of the eigenvectors for each will 
be $\vec{e}_{\phi}$. 
When integrated, this gives toroidal tendex and vortex lines (i.e., lines that 
are azimuthal, closed circles). 
The other two sets of lines for each tensor are poloidal (i.e., they lie in 
slices of constant $\phi$). 

\begin{figure*}[t!]
\includegraphics[width=0.9\textwidth]{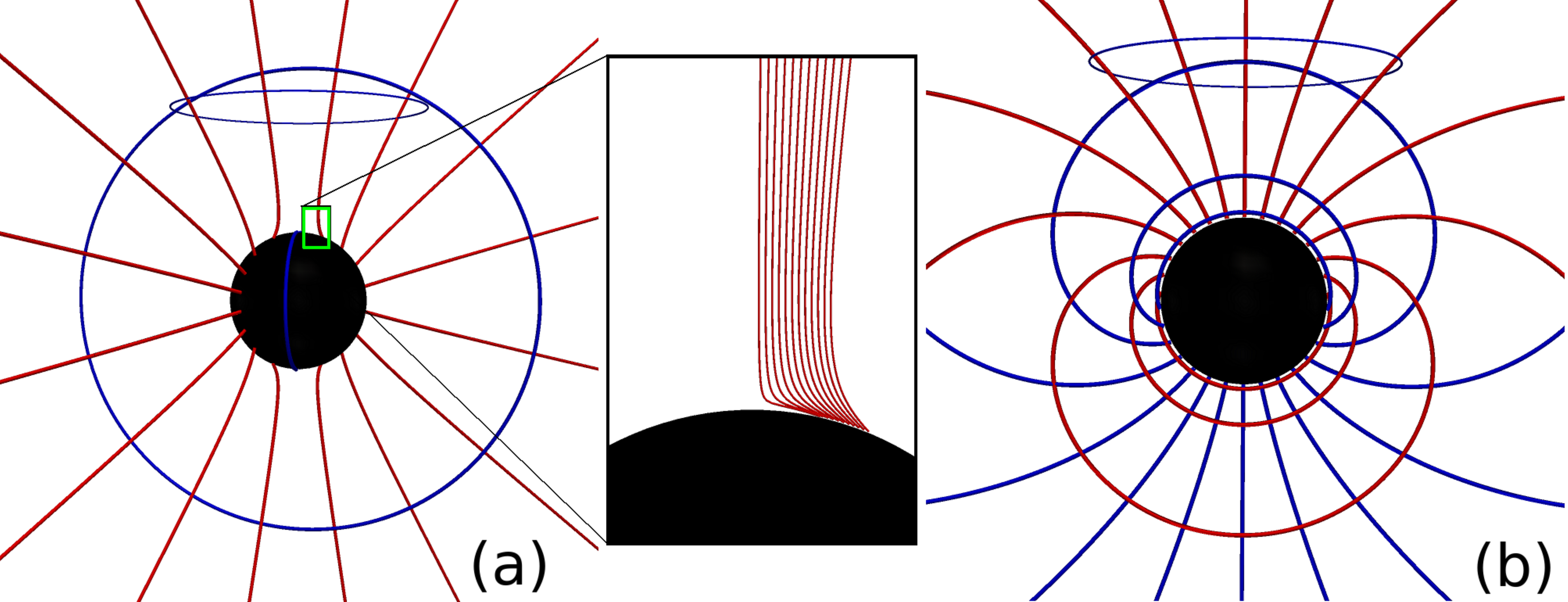}
\includegraphics[width=0.9\textwidth]{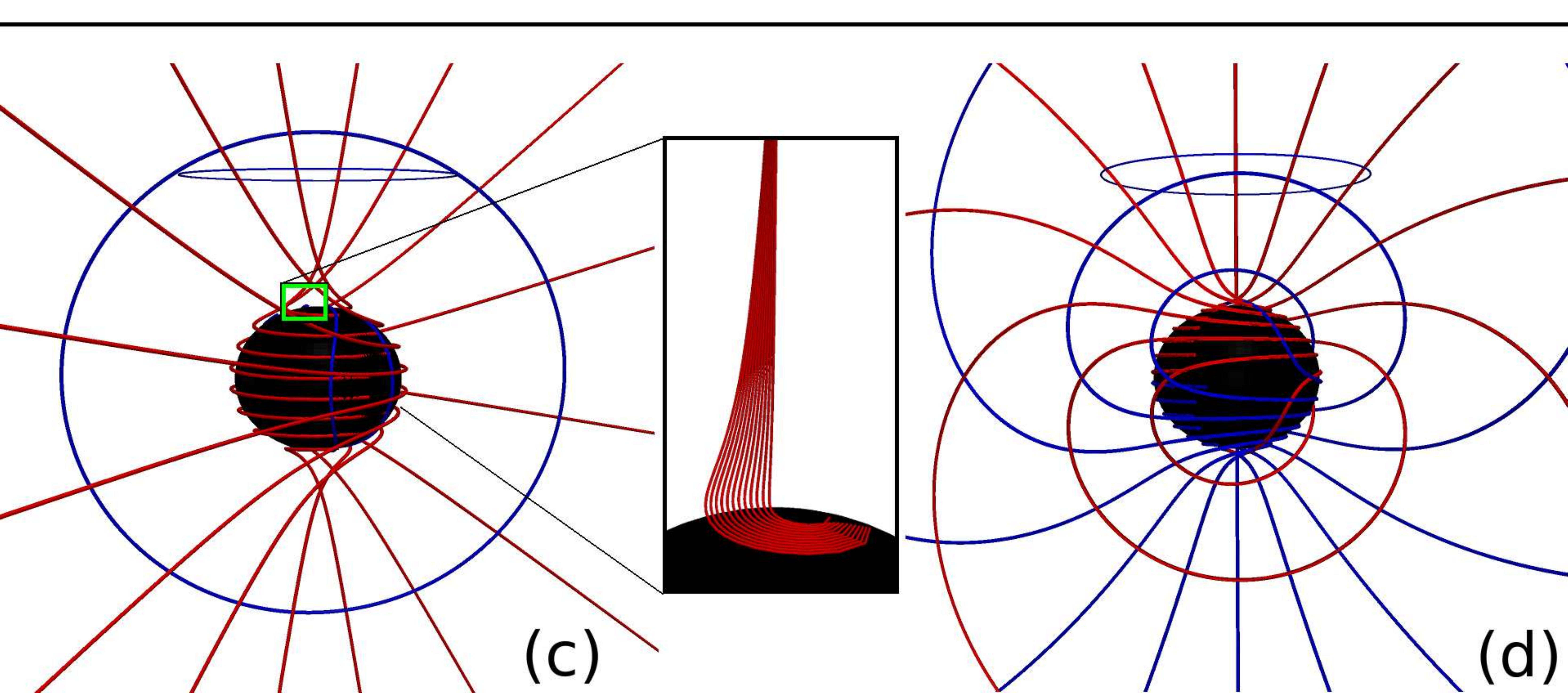}
\caption{(a) Tendex lines for a Kerr black hole with $a/M =0.875$
on a slice of constant Boyer-Lindquist time $t$, plotted in \textit{\textbf{Boyer-Lindquist spatial coordinates}}. 
The lines with positive tendicity are colored blue and negative are colored red. (b) Vortex lines for this same black hole,
slicing and coordinates, with lines of positive vorticity colored blue and negative colored red.
(c) and (d) Tendex and vortex lines for the same black hole and same Boyer-Lindquist slicing, but drawn in the 
\textit{\textbf{Kerr-Schild spatial coordinates}}. (The color of the lines and horizon are similar to Fig. 5)}
\label{fig:BL}
\end{figure*}

More specifically, the eigenvectors of the tidal field are
\ba
\vec{V}^{\mathcal{E}}_r & = & 
{
(\lambda_r^{\mathcal{E}} - \mathcal{E}_{\hat \theta \hat \theta}) \vec{e}_{\hat r} 
+  \mathcal{E}_{\hat r \hat \theta}\vec{e}_{\hat \theta} 
\over
\sqrt{(\lambda_r^{\mathcal{E}} - \mathcal{E}_{\hat \theta \hat \theta})^2 +
(\mathcal{E}_{\hat r \hat \theta})^2}
}
\,, \notag
\\
\vec{V}^{\mathcal{E}}_\theta & = & 
{
(\lambda_\theta^{\mathcal{E}} - \mathcal{E}_{\hat \theta \hat \theta}) \vec{e}_{\hat r} 
+  \mathcal{E}_{\hat r \hat \theta}\vec{e}_{\hat \theta}
\over
\sqrt{(\lambda_\theta^{\mathcal{E}} - \mathcal{E}_{\hat \theta \hat \theta})^2 +
(\mathcal{E}_{\hat r \hat \theta})^2}
}
 \,, \notag
\\
\vec{V}^{\mathcal{E}}_\phi & =& \vec{e}_{\hat \phi} \,.
\label{eq:BLev}
\ea
The labeling of these eigenvectors is such that, as $a\rightarrow 0$,
they limit to the corresponding eigenvectors (\ref{eq:SchTendicities}) 
of a Schwarzschild black hole.  
The tendicities (eigenvalues) associated with these three 
eigenvectors, which appear in the above formulas, are
\ba
\lambda_r^{\mathcal{E}} & = & -\frac{Q_e}{2} - \sqrt{\left(\frac{3Q_e}{2}\right)^2\biggl(\frac{1 + \xi}{1-\xi}\biggr)^2 + \mu^2 Q_m^2} \,, \notag
\\
\lambda_\theta^{\mathcal{E}} & = & -\frac{Q_e}{2} +  \sqrt{\left(\frac{3Q_e}{2}\right)^2\biggl(\frac{1 + \xi}{1-\xi}\biggr)^2 + \mu^2 Q_m^2} \,, \notag
\\
\lambda_\phi^{\mathcal{E}} &= & Q_e \,,
\label{BLElecLambda}
\ea
The eigenvectors of the frame-drag field are
\ba
\vec{V}^\mathcal{B}_\pm & = & 
{ (\lambda^\mathcal{B}_\pm - \mathcal{B}_{\hat \theta \hat \theta})
\vec{e}_{\hat r} +  \mathcal{B}_{\hat r \hat \theta}\vec{e}_{\hat \theta}
\over \sqrt{(\lambda^\mathcal{B}_\pm - \mathcal{B}_{\hat \theta \hat \theta})^2
+ (\mathcal{B}_{\hat r \hat \theta})^2 }
} \,, \notag
\\
\vec{V}^\mathcal{B}_\phi & = &\vec{e}_{\hat \phi} \,.
\label{eq:BLBev}
\ea
Here the labeling $+$ and $-$ of the poloidal eigenvectors corresponds to
the signs of their eigenvalues (vorticities).  The eigenvalues are
\ba
\lambda_\pm^{\mathcal{B}} &=& -\frac{Q_m}{2} \pm \sqrt{\left(\frac{3Q_m}{2}\right)^2\biggl(\frac{1 + \xi}{1-\xi}\biggr)^2 + \mu^2 Q_e^2} \,, \notag
\\
\lambda_\phi^{\mathcal{B}} &=& Q_m \,.
\label{BLMagLambda}
\ea

The tendex and vortex lines tangent to the eigenvectors
(\ref{eq:BLev}) and (\ref{eq:BLBev}) are shown in Fig.~\ref{fig:BL} 
for a rapidly rotating black hole, $a/M = 0.875$. The lines with
positive eigenvalues (tendicity or vorticity) are colored blue, and those
with negative eigenvalues are colored red. 
Far from the black hole, the tendex lines resemble those of a Schwarzschild black hole, and the vortex lines resemble those of a slowly spinning hole. However, near the horizon 
the behavior is quite different. The nearly radial tendex lines in the 
inset of Fig.\ \ref{fig:BL} are bent sharply as they near the horizon, 
because of the black hole's spin.

Before closing this appendix, we describe the behavior of the eigenvalues near 
the poles. 
From Eqs.~\eqref{BLElecLambda}, we see that as $\theta \to 0$ and 
$\theta \to \pi$, $\lambda^{\mathcal E}_\theta \to \lambda^{\mathcal E}_\phi$. 
Along the polar axis, therefore, the poloidal and axial eigenvectors of 
$\mathcal E_{\hat a \hat b}$ become degenerate, and any vector in the plane 
spanned by these directions is also an eigenvector at the axis. 
Meanwhile, for $\mathcal B_{\hat a \hat b}$, Eqs.~\eqref{BLMagLambda} show that
as $\theta \to 0$, $\lambda^{\mathcal B}_+ \to \lambda^{\mathcal B}_\phi$, and 
as $\theta \to \pi$, $\lambda^{\mathcal B}_- \to \lambda^{\mathcal B}_\phi$. 
Once again there is a degenerate plane spanned by two eigenvectors at the polar
axis. 
In Paper III, in which we study the tendex and vortex lines of perturbed Kerr 
black holes, the degenerate regions have a strong influence on the perturbed 
tendex and vortex lines (see Appendix F of Paper III).

\section{Kerr Black Hole in Kerr-Schild Slicing and Ingoing-Kerr Coordinates}
\label{app:IngoingKerr}

In ingoing-Kerr coordinates $\{\tilde t, r,\theta,\tilde\phi\}$
[Eqs.\ (\ref{eq:tKS}) and (\ref{eq:phiCS})], 
the Kerr metric takes the form (see, e.g., Chapter 33 of \cite{MTW}, though we 
use the Kerr-Schild time $\tilde t$ \cite{Kerr1963,BoyerLindquist1967}, or 
Eq.\ (D.4) of \cite{Gourgoulhon2005})
\ba
\label{KerrIngoingMetric}
ds^2 & = & - \biggl( 1 - \frac{2 M r}{\Sigma}\biggr) d\tilde t^2 + \frac{4 M r}{\Sigma} dr d\tilde t  -\frac{4 M a r \sin^2 \theta}{\Sigma} d\tilde t d \tilde \phi \nonumber \\ &&
+ H^2 dr^2 + \Sigma d \theta^2
 - 2 a H \sin^2 \theta dr d\tilde \phi + \frac{A \sin^2 \theta}{\Sigma} d\tilde \phi^2 \,, \nonumber \\
H^2 & =& 1+ \frac{2M r}{\Sigma} \,,
\ea
where $\Sigma$ and $A$ are defined in Eq.\ (\ref{eq:BLLineElem2}). 
The 4-velocities of ingoing-Kerr observers, who move orthogonally to 
slices of constant $\tilde t$, and the orthonormal tetrads they carry, are
given by 
\begin{align}
\vec{u} & = H \partial_{\tilde t} - \frac{2 M r}{H \Sigma } \partial_r \,, & \vec{e}_{\hat r} &=  \frac{\sqrt{A}}{H \Sigma} \partial_r + \frac{a H}{ \sqrt{A}} \partial_{\tilde \phi} \,, \notag \\
\vec{e}_{\hat \theta} & =  \frac{1}{\sqrt{\Sigma}} \partial_\theta \,, &
\vec{e}_{\hat {\tilde\phi}}  &=  \sqrt{\frac{\Sigma}{A}}\frac{1}{ \sin \theta} \partial_{\tilde \phi} \,
\label{eq:KerrIKtetrad}
\end{align}
(see, e.g., \cite{King1975} or \cite{Gourgoulhon2005}).

The components of the tidal field in this orthonormal basis are
\begin{widetext}
\ba
\label{EKSSlice}
\mathcal{E}_{\hat a \hat b} =\bma -Q_e \frac{2 + \xi}{1-\xi} &Q_m \frac{3 a  (r^2 +a^2) \sin \theta}{H \sqrt{A \Sigma}} & Q_e \frac{6 a M r  (r^2+a^2) \sin \theta}{H A  \sqrt{\Sigma}} \\ * & Q_e \left( 1 + \frac{3 a^2 \sin^2 \theta}{H^2 \Sigma} \right) & - Q_m\frac{6 a^2 M r \sin^2 \theta}{H^2 \Sigma \sqrt{A}} \\ *  & * &  Q_e \frac{2 + \xi}{1-\xi} - Q_e \left( 1 + \frac{3 a^2 \sin^2 \theta}{H^2 \Sigma} \right)  \ema \,,
\ea
\end{widetext}
where $Q_e$, $Q_m$, and $\xi$ are defined in Eqs.\ (\ref{eq:Qe}), 
(\ref{eq:Qm}), and (\ref{eq:Xi}).
Just as in Boyer-Lindquist slicing and coordinates (Appendix 
\ref{app:BLKerr}), so also here,
the components $\mathcal B_{\hat a \hat b}$ of the frame-drag field 
can be deduced from $\mathcal E_{\hat a \hat b}$ by the duality relation
\be
\label{BKSSlice}
\mathcal B_{\hat a \hat b} = \mathcal E_{\hat a \hat b}|_{Q_e \rightarrow
Q_m, \; Q_m \rightarrow - Q_e}\;.
\ee

The eigenvalues of the tidal field (\ref{EKSSlice}), i.e.\ the tendicities,
and their corresponding eigenvectors are
\ba
\label{KSEigenvalues}
\lambda^{\mathcal{E}}_r & = & -\frac{3 \zeta}{2 H^2 \Sigma} - \frac{Q_e}{2} \,, \notag \\
\lambda^{\mathcal{E}}_\theta & = &  \frac{3 \zeta}{2 H^2 \Sigma} - \frac{Q_e}{2} \,, \notag \\
\lambda^{\mathcal{E}}_\phi & = & Q_e \,, \\
\zeta^2 & = & Q_e^2(H^2\Sigma)^2 + \frac{(2 M a \sin \theta)^2 F}{\Sigma^3} \,,\notag \\
F & =&  r^2 + 2 M r +a^2 \,; \notag 
\ea
\begin{widetext}
\ba
\vec{V}^{\mathcal{E}}_r &=& \frac{1}{v_r} \biggl( H\sqrt{\Sigma}(r^2+a^2) \vec{e}_{\hat r} + \frac{\sqrt{A}}{2 Q_m a \sin \theta}\left[Q_e(F + a^2 \sin^2 \theta) - \zeta\right] \vec{e}_{\hat \theta}
 - 2 M a r  \sin \theta \vec{e}_{\hat {\tilde \phi}} \biggr) \,, \notag \\
\vec{V}^{\mathcal{E}}_\theta &=& \frac{1}{v_\theta} \biggl( H\sqrt{\Sigma}(r^2+a^2) \vec{e}_{\hat r} + \frac{\sqrt{A}}{2 Q_m a \sin \theta}\left[Q_e(F + a^2 \sin^2 \theta) + \zeta\right] \vec{e}_{\hat \theta}
 - 2 M a r  \sin \theta \vec{e}_{\hat {\tilde \phi}} \biggr)  \,, \notag \\ \vec{V}^{\mathcal{E}}_{\tilde \phi} & = &\frac{1}{v_{\tilde \phi}} \left(2 M a r \sin \theta \vec{e}_{\hat r} + H \sqrt{\Sigma} (r^2 + a^2) \vec{e}_{\hat \phi} \right) \,.
\label{KSEigenVecs}
\ea
\end{widetext}
Here the quantities $v_r, \, v_\theta,$ and $v_{\tilde \phi} $ are the norms of
the vectors in large parentheses (which give the eigenvectors 
$\vec V^{\mathcal E}$ unit norms).  
As for Boyer-Lindquist slicing, our $r$, $\theta$, $\phi$ 
labels for the eigenvectors and eigenvalues  
are such that as $a \to 0$, they limit to the corresponding Schwarzschild 
quantities
in Eddington-Finkelstein slicing. 
Note that although the expressions for $\vec V^{\mathcal{E}}_r$ and 
$\vec V^{\mathcal{E}}_\theta$ appear nearly identical, the coefficient of the
term in front of $\vec e_{\hat \theta}$ for $\vec V^{\mathcal{E}}_r$ includes
$-\zeta$, and that in front of $\vec e_{\hat \theta}$ for 
$\vec V^{\mathcal{E}}_\theta$ includes $+\zeta$.
This seemingly small difference determines whether the eigenvectors are 
predominantly radial or poloidal. 
Note also that the limit $a\to 0$ must be taken carefully with the vectors 
written in this form in order to recover the eigenvectors of a Schwarzschild
hole.

As for Boyer-Lindquist slicing, so also here, the 
eigenvectors and eigenvalues (vorticities) for $\mathcal B_{\hat a \hat b}$
can be derived from those for $\mathcal E_{\hat a \hat b}$ using the Kerr
duality relations:
\be
\{\vec V^{\mathcal B}_-, \vec V^{\mathcal B}_+,\vec V^{\mathcal B}_{\tilde\phi}\}
= \{\vec V^{\mathcal E}_r, \vec V^{\mathcal E}_\theta,\vec V^{\mathcal E}_{\tilde\phi}\}|_{Q_e \rightarrow Q_m,\; Q_m \rightarrow - Q_e}
\label{KSBEigenVecs}
\ee
\be
\{\lambda^{\mathcal B}_-, \lambda^{\mathcal B}_+, 
\lambda^{\mathcal B}_{\tilde\phi}\} 
= \{\lambda^{\mathcal E}_r, \lambda^{\mathcal E}_\theta,
\lambda^{\mathcal E}_{\tilde\phi}\}|_{Q_e \rightarrow Q_m\;, Q_m \rightarrow - Q_e}
\label{KSBEigenVals}
\ee
As in the case of Boyer-Lindquist slicing, so also for Kerr-Schild slicing, 
the transverse (nonradial)  
eigenvectors are degenerate on the polar axis. This can be seen, for example, 
from the form of $\mathcal E_{\hat a \hat b}$ in Eq.~\eqref{EKSSlice}, 
or from the corresponding eigenvalues in Eqs.~\eqref{KSEigenvalues}: 
as $\sin \theta \to 0$, the matrix becomes diagonal with two 
equal eigenvalues, $\lambda_\theta$ and $\lambda_\phi$. 
This is an inevitable consequence of axisymmetry.

\section{Spiraling Axial Vortex and Tendex Lines for Kerr Black Holes in 
Horizon-Penetrating Slices}
\label{app:AzimuthalSpiral}

In Figs.\ \ref{fig:CompareCoords}, \ref{fig:CompareSlicing}, and \ref{fig:BL},
the azimuthal tendex and vortex lines of a Kerr black
hole in horizon-avoiding Boyer-Lindquist slices are closed circles, while
those in horizon-penetrating Kerr-Schild and Cook-Scheel slices are outward
spirals.
In this section, we argue that outward spirals are common to a wide class of
horizon-penetrating slices, including ingoing-Kerr and Cook-Scheel slicings.
%(and speculate that they might be common to all horizon-penetrating slices).

The class of time slices that we will investigate are those that differ from 
Boyer-Lindquist slices, $t$, by a function of Boyer-Lindquist $r$,
\begin{equation}
t' = t + f(r) \, .
\end{equation}
For example, both ingoing Kerr and Cook-Scheel times fall into this category.
By computing the normal to a slice of constant $t'$ [when expressed in terms
of the locally non-rotating frame of Eq.\ (\ref{BLCoordTetrad})] we find that
\begin{equation}
\vec u ' = \sqrt{\frac{g^{tt}}{g^{t't'}}} \left(\vec u + 
\sqrt{\frac{g^{rr}}{g^{tt}}} \frac{df(r)}{dr} \vec e_{\hat r} \right) \, .
\end{equation}
Here $g^{tt}$ and $g^{rr}$ are the contravariant components of the metric in
Boyer-Lindquist coordinates, and $g^{t't'}$ are those in coordinates that use
$t'$ instead.
Defining
\begin{equation}
\gamma = \sqrt{\frac{g^{tt}}{g^{t't'}}} \, , \quad 
v = \sqrt{\frac{g^{rr}}{g^{tt}}} \frac{df(r)}{dr} \, ,
\end{equation}
we can see that the above transformation has the form of a set of local
Lorentz transformations between the locally non-rotating frame and the new
frame, and that $\gamma^2 = 1/(1-v^2)$.
This implies that we can express the timelike normal and the new radial vector 
as
\begin{subequations}
\begin{align}
\vec u' &= \gamma (\vec u + v \vec e_{\hat r}) \, , \\
\vec e_{\hat r'} &= \gamma (v \vec u + \vec e_{\hat r}) \, , 
\end{align}
\end{subequations}
and that we need not change the vectors $\vec e_{\hat \theta}$ and 
$\vec e_{\hat \phi}$ in making this transformation.

From the expressions for how the tidal and frame-drag fields transform under
changes of slicing (see Appendix B of \cite{Maartens1998b}), we find that 
we can compute the new components of the tidal field in the transformed slicing
and tetrad from the tidal and frame-drag fields in the Boyer-Lindquist slicing 
and tetrad [Eq.\ (\ref{eq:BLE}) and (\ref{eq:BLB})].
For a change in slicing corresponding to a radial boost, these general
transformation laws simplify to
\begin{subequations}
\begin{align}
\label{eq:Boost1}
\mathcal E_{\hat r'\hat r'} & = \mathcal E_{\hat r\hat r}^{\rm BL} \, ,\\
\label{eq:Boost2}
\mathcal E_{\hat r'\hat A'} & = \gamma(\mathcal E_{\hat r\hat A}^{\rm BL}
- v\epsilon_{\hat A \hat r \hat C} \mathcal B_{\hat C \hat r}^{\rm BL} ) \, ,\\
\label{eq:Boost3}
\mathcal E_{\hat A'\hat B'} & = \gamma^2[(1+v^2) 
\mathcal E_{\hat A\hat B}^{\rm BL} + v^2 \mathcal E_{\hat r\hat r}^{\rm BL}
\delta_{\hat A \hat B} - 2v\epsilon_{\hat r \hat C (\hat A} 
\mathcal B_{\hat B) \hat C}^{\rm BL} ]\,,
\end{align}
\end{subequations}
where $\hat A$, $\hat B$, and $\hat C$ $=\hat \theta$ and $\hat \phi$, and
where repeated lowered index $\hat C$ is summed over its two values.
To understand how $\boldsymbol{\mathcal B}$ is transformed, we use
the duality $\boldsymbol{\mathcal E} \rightarrow \boldsymbol{\mathcal B}$ and
$\boldsymbol{\mathcal B} \rightarrow -\boldsymbol{\mathcal E}$ in the 
transformation laws \eqref{eq:Boost1}--\eqref{eq:Boost3}.

By substituting the explicit expressions for the Boyer-Lindquist slicing and
tetrad tidal fields and the definition of $\mu$ in Eq.\ (\ref{eq:MuXi}), we
see
\begin{widetext}
\begin{equation}
\label{eq:TidalFieldBoost}
\mathcal E_{\hat a'\hat b'} = \left(\begin{array}{ccc} 
-\left(\frac{2 + \xi}{1-\xi}\right) Q_e & 
\gamma \left(\frac{3\sqrt \xi}{1 - \xi}\right) Q_m & 
\gamma v \left(\frac{3\sqrt \xi}{1 - \xi}\right) Q_e \\
* & \gamma^2 \left(\frac{1+2\xi}{1-\xi} - v^2\right) Q_e & 
- \gamma^2 v \left(\frac{3 \xi}{1-\xi}\right) Q_m \\
* & * & \gamma^2 \left(1- v^2 \frac{1+2\xi}{1-\xi}\right) Q_e 
\end{array}\right) \, .
\end{equation}
\end{widetext}
In calculating $\boldsymbol{\mathcal B}$, we could again use the duality in 
Eq.\ (\ref{BKSSlice}).

To compute the tendex lines and the tendicity, we express Eq.\ 
(\ref{eq:TidalFieldBoost}) in a new basis given by
\begin{subequations}
\label{eq:TidalBasisNew}
\begin{align}
\vec e_{\hat r''} & = \frac 1{\sqrt{1+\gamma^2 v^2 \xi}} (\vec e_{\hat r'} 
-\gamma v \sqrt \xi \vec e_{\hat \phi}) \, , \\
\vec e_{\hat \phi''} & = \frac 1{\sqrt{1+\gamma^2 v^2 \xi}} (\gamma v \sqrt\xi 
\vec e_{\hat r'} + \vec e_{\hat \phi}) \, ,
\end{align}
\end{subequations}
and where $\vec e_{\hat \theta}$ is again unchanged.
In this basis, the tidal field becomes block diagonal
\begin{equation}
\mathcal E_{\hat a''\hat b''} = \left(\begin{array}{ccc} \gamma^2\left(2v^2 - 
\frac{2+\xi}{1-\xi}\right)Q_e & \frac{3\gamma\sqrt{\xi(1+\gamma^2v^2\xi)}}
{1-\xi} Q_m & 0 \\ * & \gamma^2\left(\frac{1+2\xi}{1-\xi}-v^2\right)Q_e & 0 \\
* & * & Q_e \end{array}\right) \, .
\label{eq:TidalFieldDiag}
\end{equation}
We then see that the tendicities are 
\begin{subequations}
\label{eq:TendicitiesGen}
\begin{align}
\lambda_{r''} &= -\frac{Q_e}2 - \frac 3{2(1-\xi)} \nonumber \\
&\times \sqrt{[(1+\gamma^2 v^2\xi)+\gamma^2\xi]^2 Q_e^2
+4\gamma^2\xi(1+\gamma^2 v^2 \xi)Q_m^2} \, ,\\
\lambda_{\theta''} &= -\frac{Q_e}2 + \frac 3{2(1-\xi)} \nonumber \\
& \times \sqrt{[(1+\gamma^2 v^2\xi)+ \gamma^2\xi]^2 Q_e^2
+4\gamma^2\xi(1+\gamma^2 v^2 \xi) Q_m^2} \, ,\\
\lambda_{\phi''} &= Q_e \, ,
\end{align}
\end{subequations} 
and the corresponding vectors have an identical form to those in Eq.\ 
(\ref{eq:BLev}), when one replaces the components of the tidal field, the 
tendicities, and the unit vectors there with the equivalent (primed) quantities
in Eqs.\ (\ref{eq:TidalBasisNew})--(\ref{eq:TendicitiesGen}):
\begin{subequations}
\begin{align}
\vec{V}_{r''} & = \frac{(\lambda_{r''} 
- \mathcal{E}_{\hat \theta'' \hat \theta''}) \vec{e}_{\hat r''}
+  \mathcal{E}_{\hat r'' \hat \theta''}\vec{e}_{\hat \theta}}
{\sqrt{(\lambda_{r''} - \mathcal{E}_{\hat \theta'' \hat \theta''})^2 +
(\mathcal{E}_{\hat r'' \hat \theta''})^2}} \,, \\
\vec{V}_{\theta''} & = \frac{(\lambda_{\theta''}
- \mathcal{E}_{\hat \theta'' \hat \theta''}) \vec{e}_{\hat r''}
+  \mathcal{E}_{\hat r'' \hat \theta''}\vec{e}_{\hat \theta}}
{\sqrt{(\lambda_{\theta''} - \mathcal{E}_{\hat \theta'' \hat \theta''})^2 +
(\mathcal{E}_{\hat r'' \hat \theta''})^2}}  \,, \\
\vec{V}_{\phi''} & = \vec{e}_{\hat \phi''} \,.
\label{eq:GenEigVec}
\end{align}
\end{subequations}

From the expressions for the eigenvectors, we can explain several features of
the tendex lines in Figs.\ \ref{fig:CompareCoords}, \ref{fig:CompareSlicing}, 
and \ref{fig:BL}.
When $v=0$ [i.e., when $f(r)=0$ and the slicing is given by the 
horizon-avoiding, Boyer-Lindquist time], the azimuthal lines formed closed 
loops, and the radial and polar lines live within a plane of constant $\phi$.
For all other slicings in this family [i.e., $v\neq 0$ and $f(r)\neq 0$], the 
azimuthal lines pick up a small radial component, and they will spiral outward
on a cone of constant $\theta$ with a pitch angle whose tangent is proportional
to $\gamma v \sqrt \xi$; the radial and polar lines will also wind slightly
in the azimuthal direction (an effect that is more difficult to see in Figs.\
\ref{fig:CompareCoords} and \ref{fig:CompareSlicing}).
By duality, an identical result holds for the azimuthal vortex lines of
$\boldsymbol{\mathcal B}$, and an analogous behavior holds for the poloidal
vortex lines (in Boyer-Lindquist slicing, they remain in planes of constant
$\phi$, but in horizon-penetrating slicings, they twist azimuthally).

For this class of slices, the azimuthal eigenvector of the tidal field changes
linearly in the velocity of the boost, but the tendicity along the 
corresponding tendex line is unchanged.
The other eigenvectors also change linearly in the velocity, but their 
tendicities are quadratic in $v$; therefore, for small changes in the slicing,
the tendicities change more weakly.
This result is reminiscent of a similar qualitative result for perturbations 
of black holes in the next paper of this series: the tendex lines appear to be
more slicing dependent than their corresponding tendicities.

In the relatively general class of slicings investigated here, we showed that 
the generic behavior of the azimuthal lines in horizon-penetrating slices is to
spiral outward radially (and the other lines must also wind azimuthally as
well).
This, however, is not the most general set of slicings that still respect the 
symmetries of the Kerr spacetime [e.g., those of the form $t'=t+g(r,\theta)$
are].
These slicings will have a $\theta$ component to their boost velocities, and
(based on the argument for slowly spinning black holes in Sec.\ 
\ref{sec:SlowKerrRobustness}), the azimuthal vortex lines will also wind in 
the polar direction.
A more generic, behavior, therefore, would be azimuthal lines that no longer
wind on cones of constant $\theta$.
Because we were not aware of any simple analytical slicings of this form, we
did not investigate here; however, we suspect that this more general behavior
of the lines may appear in numerical simulations.

Before concluding, we note that by choosing
\begin{equation}
\gamma = \sqrt{\frac A{H^2\Delta\Sigma}} \,, \quad v = \frac{2Mr}{\sqrt A} \, ,
\end{equation}
we can recover the results given in Appendix \ref{app:IngoingKerr}
for the tidal field (and by duality, the frame-drag field).
Similarly, if we choose
\begin{subequations}
\begin{align}
\gamma & = \sqrt{\frac{A(r-r_-)}
{\Delta[(r-r_-)\Sigma+2M(r^2+r_+ r + r_+^2 + a^2)]}} \, , \\
v & = \frac{r_+^2 + a^2}{\sqrt A} \, ,
\end{align}
\end{subequations}
then we can use Eq.\ (\ref{eq:TidalFieldBoost}) to calculate the tidal and
frame-drag fields in time-harmonic Cook-Scheel slicing (and its associated
tetrad).
The expressions were not as simple as those in Appendix \ref{app:IngoingKerr},
and for this reason, we do not give them here.
Because the velocity in Cook-Scheel slicing falls off more rapidly in radius 
than that in ingoing-Kerr slicing the azimuthal lines should have a tighter
spiral (a feature that we observe in Fig.\ \ref{fig:CompareSlicing}).

\bibliography{References/References}

\end{document}